\documentclass[manuscript,nonacm]{acmart}
\titlenote{Preprint. Submitted to ACM Transactions on Software Engineering and Methodology (TOSEM); currently under review.}
\setcopyright{none}
\acmDOI{}
\AtBeginDocument{%
  }

\usepackage{balance}
\usepackage{enumitem}
\usepackage[T1]{fontenc}
\usepackage[utf8]{inputenc}
\usepackage{geometry}
\usepackage{hyperref}
\usepackage{booktabs}
\usepackage{tabularx}
\usepackage{array}
\usepackage{url}
\usepackage{parskip}
\usepackage{listings}
\usepackage{algorithm}
\usepackage{algpseudocode}

\usepackage{xcolor}
\definecolor{asmcomment}{RGB}{0,110,70}
\newcommand{\asmcomment}[1]{{\color{asmcomment}\ttfamily#1}}

\begin{document}

\title{MallocSan: A Memory Safety Tool for Native Closed-Source Applications}

\author{Adel Belkhiri}
\affiliation{
  \institution{Polytechnique Montreal}
  \city{Montreal}
  \state{Quebec}
  \country{Canada}}
\email{adel.belkhiri@polymtl.ca}

\author{Michel Dagenais}
\affiliation{
  \institution{Polytechnique Montreal}
  \city{Montreal}
  \state{Quebec}
  \country{Canada}}
\email{michel.dagenais@polymtl.ca}

\author{Maxime Lamothe}
\affiliation{
  \institution{Polytechnique Montreal}
  \city{Montreal}
  \state{Quebec}
  \country{Canada}}
\email{maxime.lamothe@polymtl.ca}

\renewcommand{\shortauthors}{Adel Belkhiri et al.}

\begin{CCSXML}
<ccs2012>
   <concept>
       <concept_id>10011007.10011074.10011099.10011102.10011103</concept_id>
       <concept_desc>Software and its engineering~Software testing and debugging</concept_desc>
       <concept_significance>500</concept_significance>
       </concept>
 </ccs2012>
\end{CCSXML}

\ccsdesc[500]{Software and its engineering~Software testing and debugging}

\begin{abstract}
Memory-corruption errors remain a leading cause of high-impact vulnerabilities in C and C++ software. Deterministic detection, however, remains difficult to deploy: compiler-based sanitizers require source code and control of the build pipeline, hardware-assisted schemes depend on specific platforms, and dynamic binary translation can impose order-of-magnitude slowdowns. This paper presents MallocSan, a heap sanitizer for native, potentially closed-source x86-64 Linux applications that requires no source access, recompilation, or specialized hardware. MallocSan interposes on memory allocation through \texttt{LD\_PRELOAD} and embeds an object identifier in the unused high bits of each protected pointer. Dereferencing the resulting noncanonical pointer faults at the offending instruction, which MallocSan decodes and patches at runtime so that subsequent executions perform per-object bounds checks entirely in userspace. Sites that cannot be patched fall back to in-handler emulation or single-stepping, while an optional profile-guided pass rewrites frequently executed residual sites offline. MallocSan also extends identity-based checking to vector gather/scatter instructions and supports policy-scoped coverage. Across seven SPEC CPU~2017 benchmarks, MallocSan outperforms Valgrind Memcheck on five, with a geometric-mean execution-time factor of $4.82\times$ relative to native execution, compared with $18.55\times$ for Memcheck, while preserving substantial parallel scaling on both evaluated multithreaded workloads: 644.nab\_s and pigz. On the in-scope Juliet tests, MallocSan detects all seeded violations with no false reports on the good executions. It also detects heap-buffer errors in real-world applications, including a known one-byte overread in LibTIFF's \texttt{tiffcrop} utility.
\end{abstract}

\keywords{Memory safety, heap sanitizer, buffer overflow, pointer tagging, dynamic binary patching, closed-source binaries}

\maketitle

%-----------------------

\section{Introduction}\label{sec:introduction}
C and C++ remain deeply embedded in the software infrastructure on which modern computing depends. Operating systems, language runtimes, database engines, browsers, embedded firmware, and performance-critical libraries rely on these languages because they expose a level of control that higher-level environments abstract away. However, the same control is also these languages' most persistent weakness. A pointer is not only a reference to an object but also an arithmetic value that can be copied, cast, incremented, stored, and dereferenced with limited runtime mediation. Memory correctness therefore lies largely outside the language runtime, distributed instead across programmers, libraries, compiler conventions, and testing practice. As a result, the object boundaries and lifetimes on which safe execution depends are never consistently enforced.

Memory errors arise in two closely related forms. A \emph{spatial safety} violation occurs when a program reads from or writes to an address outside the object that a pointer is intended to reference. Buffer overflows, overreads, and underflows are common examples. For instance, the expression \texttt{a[i]} compiles to pointer arithmetic with no automatic check that \texttt{i} remains within the array. A \emph{temporal safety} violation occurs when a program uses an object outside its valid lifetime. Heap use-after-free, double-free, dangling-pointer dereferences, and use-after-return errors belong to this class. Both forms emerge from ordinary idioms, including calls to \texttt{malloc} and \texttt{free}, unchecked indexing, pointer casts, and legacy string routines whose contracts depend on conventions such as NUL termination~\cite{vanoorschot2023memory,dorostkar2026addressmonitor}.

The consequences often extend beyond ordinary faults. Some invalid accesses crash immediately, whereas others corrupt adjacent state and manifest only later. An out-of-bounds write may silently damage allocator metadata or a function pointer, and a stale heap pointer may appear harmless until the freed region is reallocated. This distance between cause and symptom complicates diagnosis. Moreover, the same mechanisms can expose sensitive data, bypass program logic, or redirect control flow, which makes memory safety a prerequisite for trustworthy native software rather than a reliability concern alone~\cite{dorostkar2026addressmonitor,nsa2023memory}. Awareness and disciplined programming have not been sufficient. Major operating-system and browser vendors attribute 65--70\% of reported vulnerabilities in recent years to memory errors~\cite{vanoorschot2023memory}, and independent studies confirm that memory corruption continues to dominate high-impact failures in C and C++ software~\cite{pereira2021characterizing,oncd2024back}. This persistence is unsurprising, as the problem is rooted in the programming model itself and lies beyond the reach of coding standards or documentation, especially in systems that combine first-party and third-party code.

Memory sanitization has therefore become one of the most practical responses. A sanitizer instruments a program, its runtime environment, or its binary so that invalid accesses are detected close to the instruction that performs them, turning latent corruption into a concrete diagnostic that identifies an address, an allocation context, and a failing instruction. Developers rely on sanitizers during testing, fuzzing, debugging, and, increasingly, production-like execution. Yet the need for sanitization also reveals a deeper limitation of native code: memory safety is not a default property of the execution environment but must be reconstructed through metadata, instrumentation, runtime checks, or hardware support.

Existing approaches trade detection strength against deployability. Static and machine-learning analyzers impose no runtime overhead but warn rather than enforce, whereas compiler-based sanitizers require source code or control of the build pipeline~\cite{shahriar2010classification,serebryany2012addresssanitizer,nagarakatte2009softbound}. Hardware-assisted schemes depend on specific processors, operating-system support, ABIs, or toolchains~\cite{watson2015cheri,woodruff2019cheri,serebryany2018memory,chen2023mtsan}. Binary-level tools such as Memcheck and Dr. Memory accept unmodified executables but incur substantial dynamic-translation overhead~\cite{nethercote2007valgrind, bruening2011practical}. Runtime interposition avoids recompilation and specialized hardware, but allocator-based defenses such as DieHard and FreeGuard provide probabilistic spatial protection, while stronger pointer tracking such as CRCount still requires compiler instrumentation~\cite{berger2006diehard,silvestro2017freeguard,shin2019crcount}.

A consistent gap thus runs across these families: the most deployable tools lack deterministic identity-based spatial checking, whereas the tools that provide it require source access, recompilation, specialized hardware, or expensive whole-program translation. This gap motivates the central question of this paper: \emph{can a heap sanitizer operate on native closed-source applications on commodity x86-64 systems, without source access or recompilation, while still providing deterministic per-object spatial checking rather than probabilistic heap hardening?} We answer this question with MallocSan, a hybrid design that combines allocator interposition for ease of deployment, object-identity checking for detection accuracy, and runtime binary patching for performance.

Deployed as a shared library through \texttt{LD\_PRELOAD}, MallocSan intercepts the \texttt{malloc} family and selectively protects heap allocations. Each protected object receives an object identifier (OID), encoded in the unused high bits of the returned pointer, together with an object-table entry that records the allocation's base address, size, and allocation site. Bounds checking is thereby reduced to a single question: does the effective address derived from a pointer still fall within the object identified by its OID? Each protected dereference is checked against the recorded bounds of its corresponding allocation. As a result, even violations arising from nonlinear access patterns are reliably detected, making the detection process deterministic.

MallocSan uses the processor itself as the first interception point. A pointer carrying its high-bit tag is noncanonical on x86-64, so a direct dereference faults at the exact memory-access instruction. The runtime handles this first fault, decodes the instruction, and records it in an instruction table that later executions reuse. It then strips the OID, validates the access against the object table, executes the original operation on the canonical address, and restores the tag when the pointer must remain protected. Because purely fault-driven checking would be prohibitively expensive in hot loops, recurring checks are moved out of kernel-mediated fault handling and into userspace through runtime binary patching. Where an instruction can be patched safely, \texttt{libpatch}~\cite{dion2023libpatch} redirects the control flow to MallocSan's checking logic and to an out-of-line copy of the original instruction.

Several additional design choices make MallocSan practical for real-world binaries. Protection can be scoped by allocation size, allocation order, the number of protected objects, and caller origin, allowing overhead to be concentrated on the objects and components most relevant to an analysis. An optional guard-page backend complements OID tagging by placing selected large allocations adjacent to an inaccessible page, eliminating per-access instrumentation at the cost of additional memory and slightly reduced precision. MallocSan also extends checking to vector-indexed memory operations, validating each active lane against the bounds of the object its base pointer identifies. This capability is important because such instructions may be introduced explicitly through handwritten SIMD intrinsics or emitted automatically by the compiler under optimization and target-specific flags.

Together, these mechanisms distinguish MallocSan from the closest alternatives. It requires neither source code nor control over the build pipeline, and it replaces whole-program translation with patched user-space paths for repeated accesses. Because it reacts to code as it is reached, it needs no relocation information, at the cost of performing instruction analysis and patching at run time upon the first fault. Moreover, it runs on commodity x86-64 systems without specialized hardware. Beyond deployability, MallocSan provides deterministic per-object spatial validation for protected heap accesses and a policy-driven mode for reporting stale-pointer dereferences.

% PIGZ UPDATE: Contribution 4 distinguishes SPEC overhead from SPEC/pigz scaling.
In summary, this paper makes four contributions: \textbf{1)} MallocSan, a heap sanitizer combining allocator interposition, object-identity pointer tagging, fault-driven interception, and runtime binary patching to protect unmodified x86-64 Linux binaries; \textbf{2)} mechanisms that make identity-based checking practical for real-world machine code, including support for AVX2 gather and AVX-512 gather/scatter instructions, an extension to \texttt{libpatch} to enable the patching of instructions relocated to out-of-line buffers, and optimizations that defer or eliminate post-handlers; \textbf{3)} an optional guard-page backend that avoids per-access instrumentation for selected large allocations, and a profile-guided pipeline that converts hot trap-based sites into direct jump-based redirections offline while preserving checking semantics; and \textbf{4)} a detection-accuracy study using the heap bounds-error test cases from the NIST Juliet suite, an evaluation of execution-time and memory overhead on selected SPEC CPU~2017 workloads, and multithreaded-scaling experiments on OpenMP 644.nab\_s and the pigz compression application.

The remainder of the paper is organized as follows. Section~\ref{sec:background} reviews memory-safety violations, sanitizer detection models, and the binary patching mechanisms on which MallocSan builds. Section~\ref{sec:system-overview} gives a high-level overview of MallocSan's runtime architecture, and Section~\ref{sec:design} presents its main design components. Section~\ref{sec:implementation} describes implementation details, and Section~\ref{sec:pgo} introduces the profile-guided hot-path optimization. Section~\ref{sec:evaluation} reports the detection-accuracy and overhead evaluations, and Section~\ref{sec:discussion} discusses the implications and limitations of MallocSan. Section~\ref{sec:relwork} situates MallocSan among static analyzers, compiler-based sanitizers, hardware-assisted defenses, binary instrumentation systems, and allocator-based approaches. Finally, Section~\ref{sec:conclusion} concludes the paper by summarizing its main findings and outlining future directions.

\section{Background}\label{sec:background}
This section presents the conceptual foundations needed to understand MallocSan. The first subsection describes the memory-safety problem and the detection model that shape its goals, while the second introduces the patching techniques underlying its instrumentation layer.

% ---------------------------------------------------------------
\subsection{Memory Safety: Scope and Detection Model}\label{sec:bg-memsafety}
Memory-safety violations fall into two orthogonal classes. \emph{Spatial} safety errors occur when a pointer is used to access memory outside the bounds of its intended referent, as in a heap-buffer overflow, overread, or underflow. \emph{Temporal} safety errors occur when an object is accessed or deallocated outside its valid lifetime; common heap manifestations include use-after-free (UAF) and double-free. The present work focuses on heap-resident manifestations of both classes, which are especially difficult to protect in closed-source, prebuilt binaries.

Beyond error classification, detection mechanisms can be evaluated along four complementary dimensions. \emph{Detection timing} distinguishes immediate detection at the offending access from delayed detection after corruption has propagated, complicating root-cause attribution. \emph{Precision} concerns false positives, which reduce usability, and false negatives, which leave vulnerabilities undetected. \emph{Coverage} identifies the memory regions a technique can monitor: heap objects, stack variables, and global data present distinct recovery challenges, particularly for binary-level tools operating without source code. Finally, \emph{overhead} encompasses both execution-time cost and space consumed by structures such as shadow memory and metadata tables, determining whether a tool is practical for continuous deployment or only for test-time use.

Vintila~\emph{et al.} distinguish the \emph{conceptual detection potential} of a sanitizer from its \emph{actual implementation coverage} on a controlled test suite~\cite{vintila2025evaluating}. \emph{Location-based} approaches, such as redzones and guard pages, mark regions adjacent to allocations as poisoned or inaccessible and detect accesses that enter those regions. Their detection potential is inherently bounded because nonlinear or inter-object pointer jumps can skip the designated invalid regions. \emph{Identity-based} approaches, such as fat pointers, pointer tagging, and per-object metadata tables, associate protected pointers with records of their intended referents and validate those associations at dereference time. They offer greater conceptual detection potential but require richer metadata and systematic propagation through pointer-manipulating code. In practice, no existing design achieves complete spatial or temporal coverage, even when conceptually capable of doing so~\cite{vintila2025evaluating}. The gap between conceptual potential and realized coverage depends on both the detection strategy and the deployment model, motivating the design choices examined in the remainder of this paper.

% ---------------------------------------------------------------
\subsection{Binary and Runtime Patching}
\label{sec:bg-patching}
% ---------------------------------------------------------------

Instrumentation frameworks differ in when and at what level they interact with compiled code. Compiler-based frameworks, such as the LLVM infrastructure~\cite{lattner2004llvm} underlying AddressSanitizer (ASan)~\cite{serebryany2012addresssanitizer}, insert checks during the build process by transforming source code or an intermediate representation. This enables systematic instrumentation of selected sites but requires access to the compilation pipeline. Dynamic binary instrumentation (DBI) frameworks such as Valgrind~\cite{nethercote2007valgrind} remove the source-code requirement by translating executed code at runtime, at the cost of pervasive translation and instrumentation overhead.

Binary patching constitutes a third approach. It modifies selected machine-code instructions and redirects their execution without translating all executed code. Its runtime overhead therefore depends primarily on the execution frequency and cost of the patched sites rather than on the size of the binary. Binary patching requires neither source access nor whole-program translation. It may be applied before launch (\emph{static binary patching}) or to a running process (\emph{runtime patching}).

Static and runtime patchers share several instruction-site encoding strategies, illustrated in Fig.~\ref{fig:patch-encodings}. The \emph{TRAP} strategy replaces one byte with the \texttt{INT3} opcode (\texttt{0xCC}) and transfers control to a signal handler whenever the site executes. When sufficient space is available, the \emph{JUMP} strategy overwrites at least five bytes with a relative near jump (\texttt{jmp rel32}, opcode \texttt{0xE9}) to a trampoline, avoiding repeated signal delivery. When the target instruction is shorter than five bytes, a \emph{NOP-bridge} places a two-byte short jump at the patch site and a full five-byte jump in nearby compiler-generated alignment padding. Other techniques extend the set of sites that can use direct redirection. \emph{Instruction punning} selects the jump displacement so that its trailing bytes coincide with the opcodes of a successor instruction, while \emph{padded jumps} prepend redundant prefixes to adjust the reachable trampoline range without changing program semantics~\cite{duck2020binary,dion2023libpatch}. The displaced instruction bytes are copied to an \emph{out-of-line execution} (OLX) buffer, and any program-counter-relative operands are adjusted for the copy's new address.

\begin{figure}[t]
\centering
\begin{lstlisting}[
  basicstyle=\ttfamily\footnotesize,
  frame=single,
  xleftmargin=2em,
  escapeinside={(*@}{@*)}
]
(*@\textbf{(a) TRAP encoding}@*)
  Before:  0x1000: mov eax, [rbx+0x8]      (*@\asmcomment{; 3 bytes}@*)
  After:   0x1000: int3                    (*@\asmcomment{; 1 byte}@*)
           (*@\asmcomment{; Handler -> OLX copy -> return to 0x1003}@*)

(*@\textbf{(b) JUMP encoding (instruction >= 5 bytes)}@*)
  Before:  0x2000: mov rax, [rip+0x1234]   (*@\asmcomment{; 7 bytes}@*)
  After:   0x2000: jmp rel32 <trampoline>  (*@\asmcomment{; 5 bytes}@*)
           (*@\asmcomment{; Trampoline -> OLX copy -> return to 0x2007}@*)

(*@\textbf{(c) NOP-bridge encoding}@*)
  Before:  0x3000: xor eax, eax            (*@\asmcomment{; 2 bytes}@*)
           0x3002: ...                     (*@\asmcomment{; other code}@*)
           0x300A: nop; nop; nop; nop; nop (*@\asmcomment{; alignment pad}@*)
  After:   0x3000: jmp short 0x300A        (*@\asmcomment{; 2-byte short jump}@*)
           0x300A: jmp rel32 <trampoline>  (*@\asmcomment{; 5-byte jump}@*)
\end{lstlisting}
\caption{Core patching encodings shared by static and runtime tools. (a)~TRAP replaces one byte with \texttt{INT3}. (b)~JUMP overwrites five or more bytes with a relative near jump. (c)~NOP-bridge uses a short jump to reach nearby alignment padding, where a full five-byte jump is embedded.}
\label{fig:patch-encodings}
\Description{Three assembly listings compare patch encodings. TRAP replaces a three-byte load with a one-byte \texttt{INT3}; JUMP replaces a seven-byte load with a five-byte trampoline jump; and NOP-bridge redirects a two-byte instruction through nearby alignment padding containing the trampoline jump.}
\end{figure}

\subsubsection{Static Binary Patching}
\label{sec:bg-static-patching}

Static binary patching, also called binary rewriting, transforms a binary on disk before execution. Patch sites, redirections, and trampoline regions are determined offline, so their construction costs are paid once rather than during each run. Static rewriting is limited to code available at transformation time: shared libraries and plugins must be rewritten separately, while dynamically generated code requires a runtime mechanism.

MallocSan's profile-guided phase uses E9Patch~\cite{duck2020binary}, an x86-64 ELF rewriter that installs redirections without requiring complete control-flow recovery. E9Patch preserves original instruction addresses instead of globally relocating code and updating its incoming branches. Instructions of at least five bytes can host a direct JUMP encoding, as shown in Fig.~\ref{fig:patch-encodings}(b). Shorter sites use in-place strategies such as instruction punning, padded jumps, successor eviction, and neighbor eviction. Successor eviction redirects a following instruction to free bytes for the primary patch, whereas neighbor eviction uses nearby padding to form the NOP-bridge pattern shown in Fig.~\ref{fig:patch-encodings}(c).

\subsubsection{Runtime Patching}
\label{sec:bg-runtime-patching}

Runtime patching modifies code in the memory image of an executing process. A patcher identifies a target instruction by its virtual address, installs an appropriate redirection, and allocates reachable trampoline and OLX regions. Because patches are applied to loaded code, runtime patching can instrument the main executable, dynamically loaded libraries, and other code discovered after startup without rewriting each component in advance.

The encoding selected for a site depends on the available bytes and surrounding layout, including the length of the target instruction. A direct JUMP is preferred when the site provides at least five bytes. For shorter instructions, a NOP-bridge can use nearby alignment padding to host the full jump. When no safe jump-based encoding can be installed, the one-byte TRAP encoding provides a fallback at the cost of signal delivery on each execution. JUMP and TRAP therefore provide complementary fast and fallback paths.

MallocSan uses \texttt{libpatch} as its runtime patching substrate~\cite{dion2023libpatch}, whose authors report that it can instrument approximately 99.1\% of targeted sites across a wide range of executables. The library implements TRAP and JUMP together with in-place strategies such as NOP-bridge and instruction punning for sites that cannot host a direct five-byte jump. It allocates trampoline and OLX regions in pre-positioned executable-memory pools reachable from both position-independent executables and dynamically loaded libraries. Section~\ref{sec:design} describes how MallocSan uses these mechanisms for runtime checking, while Section~\ref{sec:pgo} describes the complementary role of E9Patch in profile-guided rewriting.

\section{MallocSan Overview}\label{sec:system-overview}

MallocSan is a heap sanitizer for \mbox{x86-64} programs. It interposes the \texttt{malloc} family via \texttt{LD\_PRELOAD} and selectively protects heap objects by encoding an \emph{object identifier} (OID) in the unused high bits of the returned pointer. Because the tagged address is noncanonical, any direct dereference triggers a hardware fault that MallocSan exploits as a precise interception point. The OID indexes an object table recording each allocation's base address and size, so every protected access can be validated against object bounds. Temporal checking is policy-driven: by default, \texttt{free} clears the table entry and recycles the OID, while an optional use-after-free mode retains the entry and marks it freed, enabling later accesses through stale pointers to be reported.

\begin{figure*}
\centerline{
\includegraphics[width=480pt]{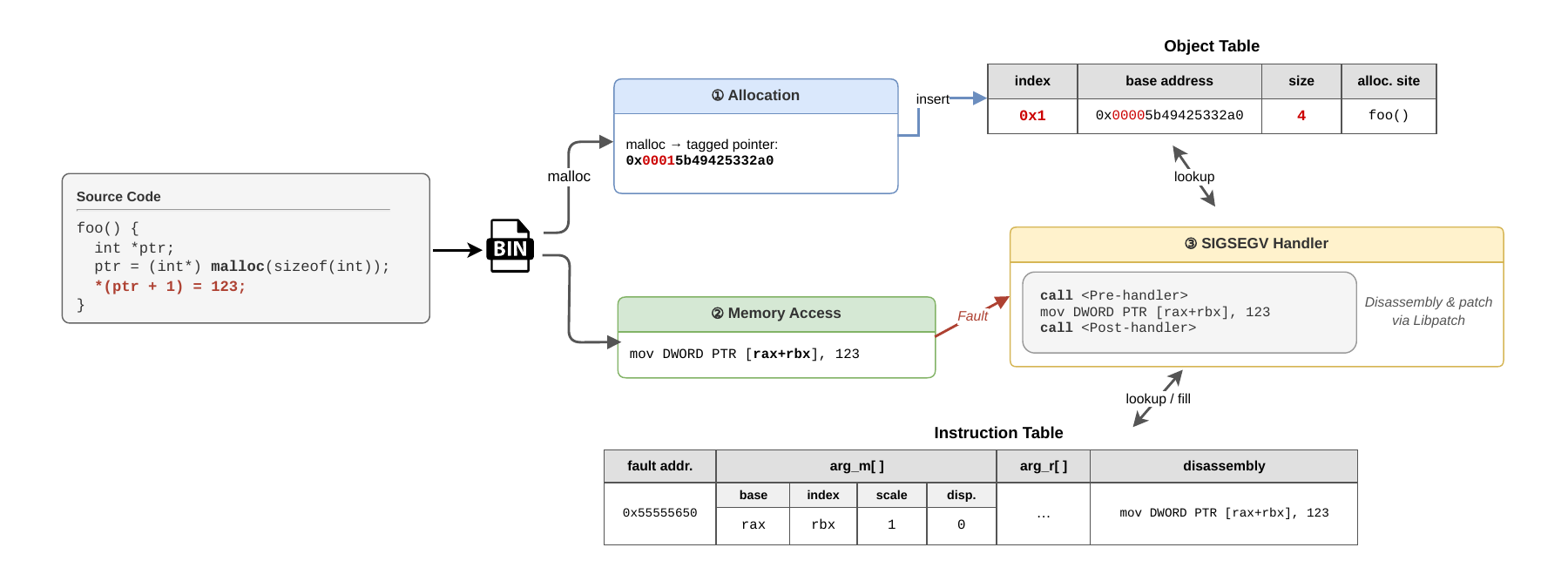}}
\vspace{-1mm}
\caption{Overview of \textsc{MallocSan}'s runtime architecture, illustrated on a heap out-of-bounds write. \label{framework-architecture}}
\Description{Flow diagram of MallocSan's runtime architecture. The interposed allocator returns a pointer tagged with an object identifier; its first dereference raises a fault, which MallocSan handles by decoding the instruction and installing a runtime patch. Subsequent executions strip the tag, validate the access against the object table's recorded base and size, execute the original instruction, and restore the tag in userspace. An out-of-bounds write is reported against the violated object's bounds.}
\vspace{-2mm}
\end{figure*}

When a tagged pointer is first dereferenced at an instruction, MallocSan captures the faulting context and decodes the instruction to recover its operand structure—including its memory and register operands—and register dependencies. If the site is patchable, MallocSan then installs a binary patch so that subsequent executions remain entirely in userspace. In the steady state, the patched path strips the OID from any tagged register before the memory access, validates the resulting address against the object table, and restores the OID so that the register again holds a protected pointer. Removing the signal from this path is what makes protection affordable because when the faulting access sits inside a hot loop, the per-iteration handler overhead would dominate execution time. If a site cannot be patched, MallocSan takes one of two slower paths. If the target instruction is simple enough, it emulates the memory access directly inside the fault handler. Otherwise it single-steps: it removes the tag in the saved machine context, sets the processor trap flag, executes the original instruction once, and restores the tag in the ensuing \texttt{SIGTRAP} handler. Such sites are rare: even instructions that \texttt{libpatch} has relocated into an out-of-line (OLX) buffer, which the earlier design could not re-patch, are usually recovered through the overlapping-patch mechanism (Section~\ref{subsec:overlapping-patches}). Fig.~\ref{framework-architecture} illustrates these mechanisms end to end on a heap out-of-bounds write. We describe these mechanisms in detail in the next sections.

\section{MallocSan Design}\label{sec:design}

MallocSan's design centers on nine components: tagged-pointer representation and object tracking, protection policy and coverage control, first-fault handling and instruction analysis, runtime patching and its fallback paths, overlapping patches for relocated instructions, post-handler elimination and deferral, C-library and kernel boundary handling, thread safety, and an optional guard-page protection backend. The following subsections address each in turn.

\subsection{Pointer Tagging and Object Table}
MallocSan tracks protected allocations at object granularity. Each protected allocation receives a nonzero OID encoded in the unused high bits of the returned pointer. The tag occupies bits~48 to~62, while bit~63 is kept clear to avoid clobbering the sign bit, yielding 15 usable tag bits and up to 32,767 distinct identifiers. The OID indexes the \emph{object table}, the authoritative record of all live protected allocations. Each entry stores the allocation base address, the object size, and the instruction pointer of the allocating call site. Bounds checking therefore reduces to an object-table lookup followed by a pointer comparison against the recorded range.

This object-identity design provides precise per-allocation bounds at every protected access and enables diagnostics to report both the faulting access site and the corresponding allocation site. The object table also determines the lifetime of each protection entry. In the default configuration, \texttt{free} clears the corresponding entry and returns its OID to the tail of the free list. A stale tagged pointer therefore no longer resolves to a live entry, and any subsequent dereference is treated as invalid unless that OID has already been reassigned.\footnote{Even after OID reuse, a stale pointer will likely fall outside the new allocation's bounds and trigger an error. An OID quarantine further reduces the chance of reassignment to an allocation containing the stale address.} When optional use-after-free tracking is enabled, \texttt{free} instead retains the entry and marks it as freed. A later dereference of the stale pointer can then be classified explicitly as a use-after-free rather than as a generic invalid access.

When all OID slots are in use, the allocator falls back to an unprotected \texttt{malloc}, emits a warning, and continues execution. By default, OIDs are recycled without a quarantine delay: upon deallocation the identifier is returned immediately to the tail of the free list via an atomic compare-exchange, making it eligible for reuse by the next protected allocation. This immediate reuse is an intentional tradeoff that keeps OID management allocation-free and lock-free. The cost is that a temporal bug can become a false negative: if a freed OID is reassigned before a stale-pointer dereference, the stale pointer resolves to the new object's entry and may pass validation if the accessed range falls within that object's bounds. Enabling use-after-free tracking avoids this immediate-reuse ambiguity for freed objects retained in the table, at the cost of increasing pressure on the finite OID space.

\subsection{Protection Coverage Control}
MallocSan stays practical on memory-intensive applications because it tags only the pointers returned by selected allocation sites: overhead tracks the scope of protection, not total heap activity. The runtime can limit protection according to allocation size, allocation order, and the maximum number of concurrently protected objects. Unprotected allocations follow the original \texttt{malloc} path without modification, while protected allocations retain their tagged representation in memory and registers throughout ordinary execution.

A complementary caller-origin filter further refines this selection. At each interposed \texttt{malloc} call, the runtime captures the caller's return address and compares it against a table that separates the main executable's address range from dynamically loaded libraries. Allocations originating from the main object and allowlisted libraries receive tags; those from other library code do not. This approach reduces overhead in workloads where most heap activity is library-internal and avoids spurious faults in library code that relies on pointer arithmetic incompatible with the tagged representation. The current implementation uses a configurable allowlist of shared objects, providing fine-grained coverage control independent of address-space layout.

\subsection{First-Fault Handling and Instruction Analysis}
Dereferencing a tagged pointer produces a synchronous fault at the offending instruction. MallocSan intercepts \texttt{SIGSEGV} and \texttt{SIGBUS} on a dedicated alternate signal stack and retains control of \texttt{SIGTRAP} to implement single-step fallbacks and to cooperate with \texttt{libpatch}. To prevent its own handlers from being overridden, MallocSan saves any application-installed signal handlers for \texttt{SIGSEGV}, \texttt{SIGBUS}, and \texttt{SIGTRAP}, ensuring that segmentation faults unrelated to tagged pointers are correctly forwarded to the application's handlers.

For instructions not yet patched, the segmentation fault handler records the machine context and fault metadata, then redirects execution to a trampoline. This indirection is necessary because Capstone, the disassembly library used for instruction analysis, is not async-signal-safe and cannot be invoked directly from a signal handler. The handler is therefore deliberately narrow: it performs only minimal bookkeeping and defers all decoding and patching decisions to the trampoline context. The trampoline preserves general-purpose registers, flags, and floating-point or vector state before invoking runtime code that can safely analyze the faulting instruction. This separation is essential for correctness because first-fault processing may require disassembly, instruction-table updates, and interactions with patch-management code, none of which can safely execute inside the handler.

MallocSan decodes the faulting instruction fully to prevent repeated faults on the same memory access. Capstone identifies memory operands, resolves base and index registers, records register read and write sets, determines the access width, detects repeat prefixes, and classifies special cases such as vector-indexed accesses. The result is an \emph{instruction-table entry} that caches everything needed to handle later executions of the same site. The instruction table is a hash table keyed by instruction address, with explicit entry states (empty, initializing, ready, and failed) that ensure at most one thread analyzes a newly faulting instruction while all others observe a stable result.

Two instruction classes require specialized analysis and bounds-validation logic. \texttt{REP}-prefixed string instructions access a multi-element range over repeated iterations, whereas vector-indexed instructions use vector indices to generate a distinct, potentially noncontiguous address for each active lane. The following subsections explain how MallocSan derives and validates these memory footprints.

\subsubsection{REP String Instructions}\label{subsec:rep-strings}

In \mbox{x86-64}, string instructions such as \texttt{MOVS}, \texttt{STOS}, \texttt{LODS}, \texttt{CMPS}, and \texttt{SCAS} may carry a repeat prefix. The prefix causes the processor to execute the elementary operation repeatedly while updating the relevant implicit pointer registers—\texttt{rsi}, \texttt{rdi}, or both—according to the direction flag (\texttt{DF}). The initial value of \texttt{rcx} specifies the maximum number of iterations. For fixed-count operations such as \texttt{REP MOVS} and \texttt{REP STOS}, it also determines the exact number of iterations. For data-dependent operations such as \texttt{REPE}/\texttt{REPNE CMPS} and \texttt{SCAS}, however, a comparison may terminate execution early, so \texttt{rcx} provides only an upper bound.

Compilers routinely emit these instructions for \texttt{memcpy}-, \texttt{memset}-, and \texttt{strlen}-style operations. Unlike an ordinary instruction, whose memory operands have fixed widths, a repeat-prefixed instruction may access a range whose extent is known only at run time. Instruction analysis therefore marks every repeat-prefixed site in its instruction-table entry and separately identifies the data-dependent forms, which require a different validation strategy.

MallocSan validates fixed-count repetitions entirely before execution. Given an untagged initial address $A$, an element size $s$, and an iteration count $n=\texttt{rcx}$, the pre-handler derives the complete memory footprint and checks it against the referenced object's bounds. For $n>0$ and $\texttt{DF}=0$, the footprint is $[A,\ A+n \times s)$. For $\texttt{DF}=1$, it is
$[A-(n-1)\times s,\ A+s)$. When $n=0$, the instruction performs no memory access and requires no bounds check. The span computation saturates on multiplication overflow, ensuring that an oversized count is reported as out of bounds rather than wrapping into an apparently valid range. Instructions with multiple memory operands are handled independently: for \texttt{MOVS}, for example, the source and destination footprints are each checked against the object referenced by their respective tagged pointers. Fig.~\ref{fig:rep-copy} illustrates this pre-execution validation for a \texttt{memcpy}-style copy.

\begin{figure}[t]
\centering
\begin{lstlisting}[
  basicstyle=\ttfamily\footnotesize,
  frame=single,
  xleftmargin=2em,
  escapeinside={(*@}{@*)}
]
(*@\asmcomment{; rsi = tagged pointer to a protected source}@*)
(*@\asmcomment{; rdi = tagged pointer to a protected destination}@*)
mov   rcx, 64     (*@\asmcomment{; iteration count: exactly 64}@*)
rep   movsb       (*@\asmcomment{; copy 64 bytes from [rsi] to [rdi]}@*)
\end{lstlisting}
\caption{A fixed-count \texttt{REP} copy implementing a 64-byte \texttt{memcpy}. Because \texttt{rep movsb} executes exactly \texttt{rcx} iterations, the pre-handler derives the complete source and destination footprints before execution and validates each interval against the bounds of the object referenced by its corresponding tagged pointer.}
\label{fig:rep-copy}
\Description{Assembly listing with tagged source register \texttt{rsi} and destination register \texttt{rdi}. It sets \texttt{rcx} to 64 and executes \texttt{rep movsb}, causing MallocSan to validate both 64-byte ranges before the copy.}
\end{figure}

Data-dependent repetitions cannot be validated from \texttt{rcx} alone. In the \texttt{strlen} idiom shown in Fig.~\ref{fig:rep-scan}, \texttt{repne scasb} begins with \texttt{rcx} set to the sentinel value $2^{64}-1$ but terminates when it encounters the first NUL byte. The actual extent therefore depends on memory contents, and treating \texttt{rcx} as the exact iteration count would incorrectly classify nearly every such scan as out of bounds. MallocSan instead divides validation between the pre-handler and the post-handler. If \texttt{rcx} is nonzero, the pre-handler validates the first element—the only access known to occur before the result of the comparison is available—and then allows the instruction to execute with untagged pointer registers. After every completed iteration, the processor adjusts each relevant pointer by one element in the direction specified by \texttt{DF}. The final pointer is therefore one element beyond the last access in the direction of traversal. By comparing the initial and final pointer values, the post-handler can recover the exact interval accessed, including its direction.

While the register still contains its untagged final value, the post-handler combines the saved OID with the lower endpoint of the recovered interval. This reconstructed tagged address allows the bounds check to resolve the appropriate object-table entry. After validating the interval, the post-handler restores the tag to the updated pointer register. This deferred validation detects scans that run beyond an unterminated buffer. Although it is a deliberate exception to MallocSan's check-before-execute rule, it allows data-dependent repetitions to be checked using their actual extents rather than potentially enormous upper bounds.

\begin{figure}[t]
\centering
\begin{lstlisting}[
  basicstyle=\ttfamily\footnotesize,
  frame=single,
  xleftmargin=2em,
  escapeinside={(*@}{@*)},
]
(*@\asmcomment{; rdi = tagged pointer to a protected buffer}@*)
xor   al, al      (*@\asmcomment{; scan value: the NUL terminator}@*)
mov   rcx, -1     (*@\asmcomment{; iteration limit: effectively unbounded}@*)
repne scasb       (*@\asmcomment{; scan byte-wise until [rdi] matches al}@*)
\end{lstlisting}
\caption{A data-dependent \texttt{REP} scan implementing \texttt{strlen}. Because \texttt{rcx} provides only an upper bound on the iteration count, the pre-handler validates only the first byte before execution. After execution, the post-handler derives the actual scanned range from the change in \texttt{rdi} and validates it against the bounds of the object referenced by the original tagged pointer.}
\label{fig:rep-scan}
\Description{Assembly listing with tagged buffer register \texttt{rdi}. It clears \texttt{al}, sets \texttt{rcx} to minus one, and executes \texttt{repne scasb}; MallocSan checks the first byte before the scan and derives the complete scanned range afterward.}
\end{figure}

\subsubsection{VSIB Instructions}
In addition to scalar memory accesses, MallocSan supports instructions that use the \emph{Vector Scaled Index Byte} (VSIB) addressing mode. These instructions, enabled by AVX extensions, may appear in compiler-generated code (e.g., through auto-vectorization with flags such as \texttt{-O3} and \texttt{-march=native}) as well as in handwritten SIMD intrinsics. Supporting them is therefore essential for extending memory safety from scalar to vector memory operations without constraining compiler optimization settings.

VSIB instructions perform multiple lane-wise loads or stores from independently computed addresses. The effective address for lane~$i$ is $\mathit{base} + \mathit{vreg}[i] \times \mathit{scale} + \mathit{disp}$, where the base is a general-purpose register and the vector index register provides per-lane integer offsets \footnote{In some cases, the base is zero, and each index is itself a complete address that may be tainted.}. AVX2 gather instructions use an explicit vector mask operand, whereas AVX-512 gather and scatter instructions use dedicated mask registers (e.g., \texttt{k1}--\texttt{k7}). In both cases, inactive lanes must neither be accessed nor subjected to bounds checks.

\begin{figure*}[t]
\centering
\includegraphics[width=1\textwidth]{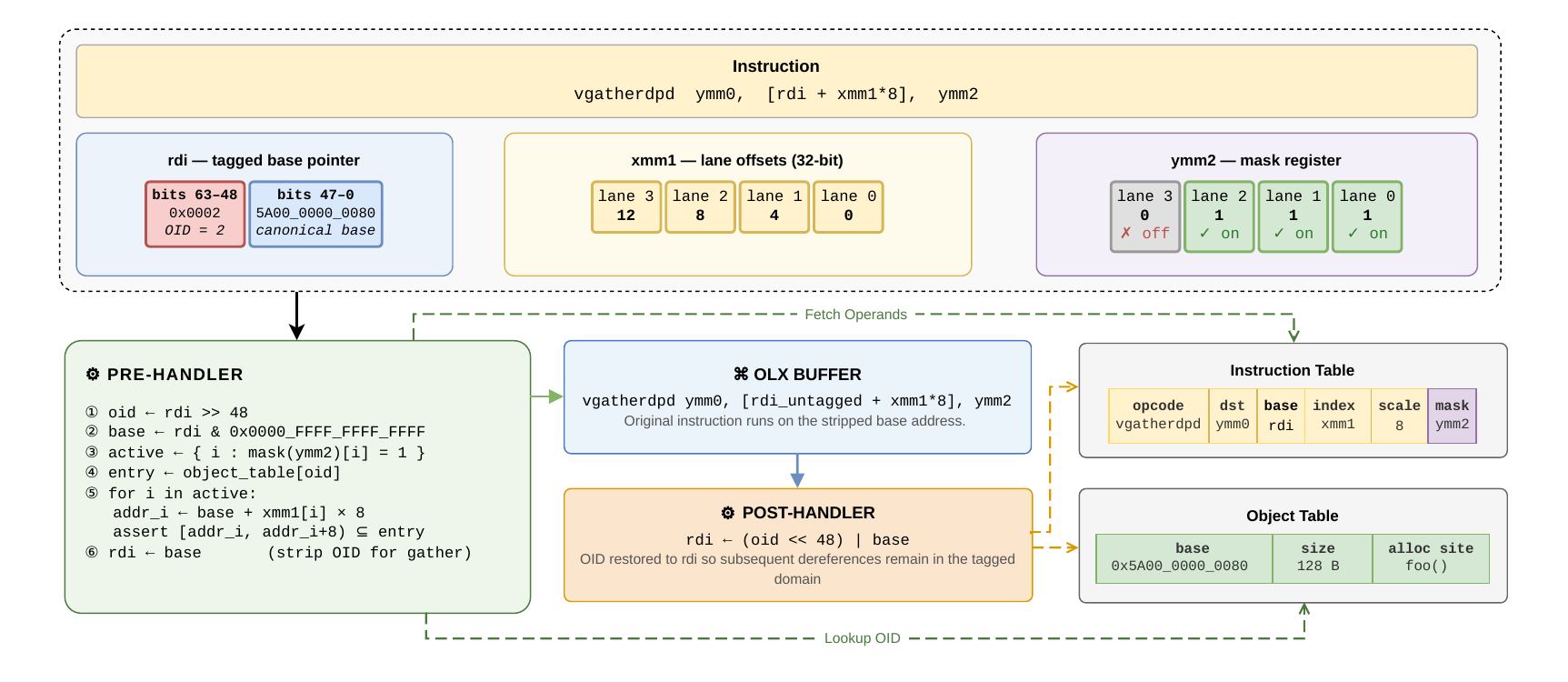}
\caption{AVX2 VSIB gather checking in MallocSan. For a four-lane \texttt{vgatherdpd}, the pre-handler extracts the OID from tagged \texttt{rdi}, restores the canonical base address, derives the active lanes from \texttt{ymm2}, and checks each enabled \texttt{xmm1}-indexed lane against the bounds of the referenced object. The gather then executes on the untagged base, and the post-handler re-tags \texttt{rdi}.}
\label{fig:vsib-gather}
\Description{Flow diagram of MallocSan handling an AVX2 \texttt{vgatherdpd} instruction. Tagged base register \texttt{rdi}, index vector \texttt{xmm1}, and mask vector \texttt{ymm2} enter a pre-handler that extracts the object identifier from \texttt{rdi}, restores the canonical base address, derives the active lanes from \texttt{ymm2}, and checks each enabled lane's effective address against the referenced object's bounds while skipping masked-off lanes. The gather then executes on the untagged base, and a post-handler restores the tag to \texttt{rdi}.}
\end{figure*}

VSIB accesses differ from scalar ones in that a single instruction may touch several lane-specific addresses. MallocSan therefore records static VSIB metadata in the instruction-table entry, including the index-element width, the lane count, and the architectural registers supplying the base, index, and mask operands. At execution time, the pre-handler reads the current index-vector contents and active-lane mask from the saved register state, strips the OID from the base register, and checks each active lane individually against the object-table entry referenced by that OID (Fig.~\ref{fig:vsib-gather}). If any lane violates the recorded bounds, the report identifies both the lane number and the computed address. When the base register carries a tagged pointer, the post-handler restores its OID if the pointer value remains live after the instruction. In this base-tagged form, the index vector carries integer offsets rather than pointers and is therefore never tagged.

\subsection{Runtime Patching and Fallback Paths}\label{subsec:patching-fallbacks}

MallocSan leverages \texttt{libpatch} to atomically rewrite the faulting instruction, redirecting control to handler code and an out-of-line (OLX) buffer where the relocated instruction is subsequently executed. The ideal case is the \texttt{JUMP} strategy, which overwrites the target site with a 5-byte \texttt{JMP} and therefore applies only when the instruction is at least five bytes long. Because \texttt{libpatch} is not thread-safe, MallocSan delegates patch installation to a dedicated worker thread once the process becomes multithreaded.

After the first fault, installed JUMP-based patches keep subsequent executions entirely in userspace, eliminating repeated fault handling. A patch diverts execution through a pre-handler that strips OIDs from the relevant base and index registers, reconstructs the effective address, and validates the access against the object table. The original instruction then executes on untagged addresses. If the instruction logically preserves a pointer-bearing register, a post-handler reapplies the saved OID to that register's updated value. This restoration is a correctness requirement: without it, subsequent instructions using the same register as a pointer operand would escape the tagged domain and bypass bounds checking entirely.

The pre/post-handler split is necessary because x86 instructions can simultaneously consume and transform the same registers. MallocSan records the removed OIDs and any auxiliary repair state before the instruction executes, and explicitly handles the case in which a register serves both as a memory-address component and as an explicit write destination, ensuring that transient untagging does not leak into program-visible state.

This ideal scenario is not always available. For instructions shorter than five bytes, \texttt{libpatch} may still construct a \texttt{JUMP}-class redirect through instruction punning or a NOP bridge (Section~\ref{sec:bg-patching}). Both tactics consume a contiguous byte span at the target site, which \texttt{libpatch} relocates into the OLX buffer. If the relocated sequence contains additional memory-accessing instructions, they cannot be patched independently at their original addresses. Section~\ref{subsec:overlapping-patches} describes our extension to libpatch for supporting overlapping patches, which enables MallocSan to instrument these instructions as well.

When no \texttt{JUMP}-class patch can be installed, MallocSan provides three fallback paths, ordered from fastest to slowest. The fastest is \emph {in-handler emulation}. For a supported instruction whose semantics can be reproduced directly from its decoded metadata and the saved machine context, the \texttt{SIGSEGV} handler computes the untagged effective address, validates it against the object table, performs the load or store, and advances the instruction pointer past the instruction. The access therefore requires only one signal delivery. Moreover, because the handler never strips tags from the architectural address registers, the protection model remains unchanged.

The current implementation supports plain \texttt{MOV} instructions that address memory using a single general-purpose base register and a constant displacement, without an index register, and that access between one and eight bytes. In our measurements, this path is approximately 1.6$\times$ as fast as a \texttt{TRAP} patch. Because validation precedes emulation, MallocSan never emulates an out-of-bounds or stale-pointer dereference.

Instructions too complex for in-handler emulation are handled using the \texttt{TRAP} strategy. When post-execution processing is required, the trap-patched instruction may be paired with a subsequent \texttt{JUMP}-patched site that hosts the deferred post-handler (Section~\ref{subsec:deferral}). Although this strategy is almost always feasible, it incurs signal delivery and handler dispatch every time the instruction executes.

\begin{figure*}[t]
\centering
\includegraphics[width=0.9\textwidth]{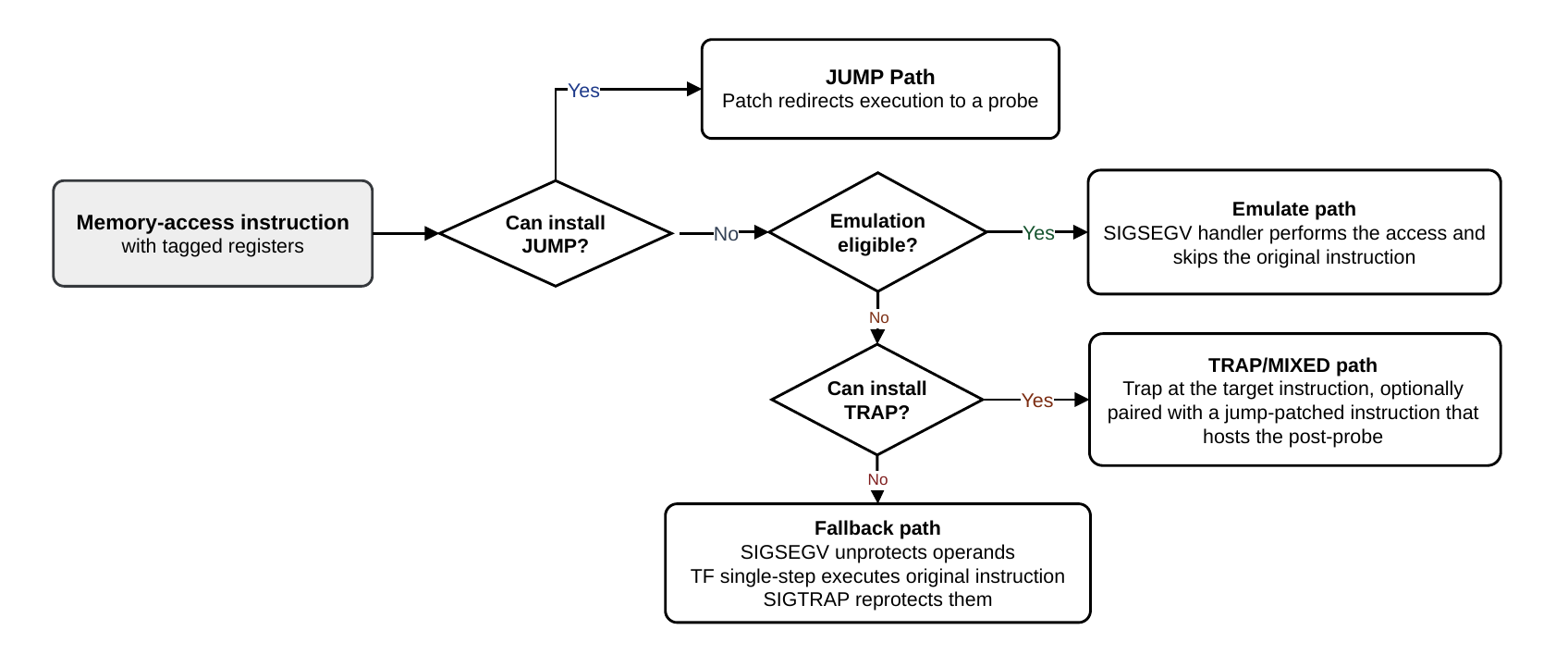}
\caption{MallocSan's patching decision logic: a \texttt{JUMP} patch is attempted first, then in-handler emulation, then a \texttt{TRAP} patch, with trap-flag single-stepping as the last resort.}
\label{fig:patching-logic}
\Description{Decision flowchart for handling a memory-access instruction whose registers hold tagged pointers. If a JUMP patch can be installed, the patch redirects execution to a probe. Otherwise, if the instruction is eligible for emulation, the SIGSEGV handler performs the access itself and skips the original instruction. Otherwise, if a TRAP patch can be installed, the site traps at the target instruction, optionally paired with a jump-patched instruction that hosts the post-handler. If none of these applies, the SIGSEGV handler strips the tags from the operands, a trap-flag single step executes the original instruction, and the SIGTRAP handler restores the tags.}
\end{figure*}

The last resort, used when a site can be neither patched nor emulated, is \emph{single-stepping}. The \texttt{SIGSEGV} handler extracts and saves the OIDs from the relevant registers, validates the access, and, if validation succeeds, writes the corresponding untagged addresses to the saved machine context. It then sets the processor's trap flag and resumes execution, causing the original instruction to execute exactly once. The resulting \texttt{SIGTRAP} handler restores the saved OIDs to the relevant registers, clears the trap flag, and resumes normal execution. Fig.~\ref{fig:patching-logic} summarizes this patching and fallback hierarchy. In practice, the vast majority of sites receive \texttt{JUMP} patches, while only a small fraction require emulation, \texttt{TRAP} patches, or single-stepping.

Although MallocSan targets binaries that cannot be rebuilt, users who can recompile an application can reduce the fallback rate further by enabling \texttt{-fpatchable-function-entry=5,5}. This option places five \texttt{NOP}s immediately before each function's entry point. Because normal control transfers enter the function after this padding, \texttt{libpatch} can repurpose the additional space without overwriting application instructions. These spare bytes facilitate \texttt{JUMP}-based redirects for otherwise too-short instructions near function entries, measurably reducing reliance on the fallback paths.

\subsection{Overlapping Patches for Relocated Instructions}\label{subsec:overlapping-patches}
Installing a patch can inadvertently make neighboring instructions unpatchable. To insert a 5-byte \texttt{JMP}, \texttt{libpatch} must overwrite enough complete instructions to cover at least five bytes. It relocates the displaced instructions to the patch's OLX buffer, where they execute before control returns to the original code. If a relocated instruction subsequently dereferences a tagged pointer, the resulting fault occurs at its copy in the OLX buffer rather than at its original address. Because patches cannot be installed inside an OLX buffer and the original instruction is already covered by the existing patch, the site cannot be patched directly. Without overlapping-patch support, every subsequent execution of that instruction would therefore require single-stepping—an especially costly fallback when the instruction lies in a hot loop.

To remove this restriction, we extended \texttt{libpatch} with \emph{overlapping patches}. The mechanism builds on the observation that the trampoline reaches the OLX buffer through an indirect pointer, which provides a natural update point. A patch requested on a relocated instruction is therefore registered as a \emph{child} of the parent patch that owns the relocated region, and no new trampoline is created. Instead, \texttt{libpatch} rebuilds the parent's OLX contents into a fresh buffer, this time surrounding the target instruction with the child's pre-handler and, when tag restoration is required, its post-handler. The new buffer is then published atomically through the parent's indirect pointer, and the old buffer is reclaimed lazily. A concurrent thread executing during the handoff observes either the old or the new buffer, both of which remain mapped and valid, so the transition requires no stop-the-world synchronization.

Overlap children follow stricter rules than ordinary patches. Because a child shares the parent's patch handle, it has no independent removal path: dropping the handle would tear down the parent and every sibling child. MallocSan therefore either commits a child probe or degrades without dropping it. Children are also excluded from post-handler deferral and restore tags immediately after the instruction. Finally, only the displaced non-head instructions of a relocated region are eligible to become children. When a child cannot be created, the site falls back to the single-step path.

\subsection{Post-Handler Elimination and Deferral}\label{subsec:deferral}

At frequently executed sites, steady-state overhead is dominated by recurring pre-handler and post-handler execution rather than by one-time first-fault analysis. This cost is particularly high inside loops, where bounds validation, tag removal, and tag restoration may recur on every iteration. MallocSan therefore applies two complementary optimizations to the post-handler path: elimination when restoration is unnecessary and deferral when restoration can safely occur later.

First, MallocSan eliminates the post-handler when the original values of all tagged registers become dead, either because the faulting instruction overwrites them or because a subsequent instruction overwrites them before any further pointer use. In such cases, restoring the saved OIDs would be unnecessary or incorrect. The pre-handler still removes the tags and validates the access, but no post-handler is installed. Fig.~\ref{fig:corner-case-overwrite} illustrates this case.

\begin{figure}[t]
\centering
\begin{lstlisting}[
  basicstyle=\ttfamily\footnotesize,
  frame=single,
  numbers=left,
  xleftmargin=2em,
  escapeinside={(*@}{@*)}
]
mov rax, qword ptr [rax + rbx*8]
\end{lstlisting}
\caption{Post-handler elimination when \texttt{rax} serves as both the memory-address base and the load destination. Because the load unconditionally replaces \texttt{rax}, reapplying its original OID would incorrectly tag the loaded value. MallocSan therefore omits the post-handler.}
\label{fig:corner-case-overwrite}
\Description{Single-instruction assembly listing of \texttt{mov rax, qword ptr [rax + rbx*8]}. The instruction uses \texttt{rax} to form the memory address and then replaces it with the loaded value, so MallocSan does not restore the original pointer tag.}
\end{figure}

Second, when restoration remains necessary but need not occur immediately, MallocSan defers the post-handler to a later safe site. The runtime obtains the address range of the containing function through \texttt{libunwind} and uses Capstone to scan subsequent instructions. The scan stops at control-flow transfers and at instructions whose interaction with the tagged registers cannot be modeled safely. A full overwrite instead eliminates the need for restoration. The principal control-flow exception, described below, is a conditional back-edge that encloses the originating instruction. Algorithm~\ref{alg:post-handler-deferral} summarizes the scan used to select a restoration site and determine whether a post-handler remains necessary.

The scanner advances the candidate restoration site only after encountering another fault-like access beyond the current candidate. This requirement ensures that each advancement meaningfully extends the untagged execution window. Because the preferred candidate may not have enough space for a 5-byte \texttt{JMP}, the scanner also retains a bounded set of later tail-fallback candidates. If no candidate can host the deferred post-handler, restoration remains immediately after the originating instruction.

Four rules govern the forward scan. An instruction is \emph{tag-independent} if it neither reads nor writes a register in the tagged set and does not reference one in a memory operand. The scanner may pass over such an instruction. However, the instruction may host the deferred post-handler only if it cannot itself fault on a tagged heap pointer. This condition holds when it has no memory operand; is a \texttt{LEA} or multi-byte \texttt{NOP}, neither of which dereferences its operands; or uses only non-heap addressing, such as FS/GS-segment-relative, RIP-relative, or \texttt{rsp}-based addressing. MallocSan deliberately does not trust \texttt{rbp}, because frame-pointer omission allows it to hold an arbitrary pointer.

A memory access is \emph{fault-like} if it uses the same tagged base or index register as the originating instruction. During the scan, MallocSan recomputes the effective address of an access whose scale or displacement differs from the original access and verifies that it lies within the same object bounds. If accepted, the access is incorporated into the widened footprint validated by the originating pre-handler. It therefore remains covered even though the tagged register is untagged by the time the instruction executes. Each accepted fault-like access advances the deferral window.

A \emph{constant pointer walk} is an \texttt{add}, \texttt{sub}, \texttt{inc}, or \texttt{dec} that adjusts a tagged 64-bit register by an immediate value. Such an instruction does not terminate the scan. Instead, MallocSan folds the adjustment into a per-register delta and validates subsequent fault-like accesses against the correspondingly shifted address. A cap on the accumulated magnitude prevents the delta from reaching the tag field.

During the forward scan, an instruction \emph{fully overwrites} the tagged set if it unconditionally replaces every remaining tagged register with a value independent of the original pointer, without first using that value as a pointer input. Once this occurs, restoration is unnecessary and the post-handler can be eliminated. MallocSan recognizes register-zeroing idioms such as \texttt{xor}~$r,r$ as a special case, even though their decoded register-access sets may syntactically include a read of the destination.

The forward scan constructs a deferral plan, but memory safety ultimately depends on a runtime invariant. The scan initially reasons using the register values observed at the first fault, whereas subsequent executions may begin with different values. On every execution, the pre-handler at the originating site therefore validates a \emph{widened footprint} against the object bounds using the tagged address register's current value. The scan records this footprint as the displacement interval spanning the originating access and every accepted fault-like access before the restoration point, including any accumulated pointer-walk deltas. All accesses within the deferred window are consequently covered before the tag is removed. If the widened footprint exceeds the object bounds, MallocSan reports the violation at the originating site before entering the window.

A loop back-edge would ordinarily terminate the scan. At a site that remains on the fault-driven path, restoring the tag before the back-edge would cause the protected access to fault again on every iteration. MallocSan avoids this per-iteration cost for a common straight-line, single-exit loop shape. When the scanner encounters a conditional backward branch whose target precedes the originating instruction, it treats the branch as a candidate loop latch and its fall-through as a candidate exit. Deferral may cross the back-edge only if all of the following conditions hold: \textbf{1)} The region from the originating instruction to the latch contains no other control-flow transfer; \textbf{2)} The loop prefix from the branch target to the originating instruction neither transfers control nor reads, writes, or dereferences through a tagged register; \textbf{3)} The protected instruction has a single non-indexed memory operand whose effective address is formed from a loop-invariant base; \textbf{4)} The latch does not access the tagged register; and \textbf{5)} A provably non-heap restoration site exists within a bounded distance after the fall-through.

For an eligible fault-driven site, the fault on loop entry invokes the pre-handler, which validates the loop-invariant address and removes its tag. Because the address does not change, this single check covers every iteration. The loop then executes natively with the untagged pointer, and one post-handler at the exit restores the tag. Multiple protected sites in the same loop can share the exit post-handler through a small registry. This optimization therefore converts a per-iteration signal cost into a once-per-loop-execution cost.

The repeat-prefixed string instructions described in Section~\ref{subsec:rep-strings} are excluded from both optimizations. Capstone reports the relevant implicit pointer registers---\texttt{rsi}, \texttt{rdi}, or both---as written because the processor advances them after each iteration. These writes do not kill the pointers: their updated values may still reference the protected objects and must therefore be re-tagged. MallocSan consequently overrides the generic write-set classification and restores their tags immediately after the instruction. Data-dependent \texttt{CMPS} and \texttt{SCAS} operations impose an additional constraint because their exact footprints can be established only after execution. Their post-execution checks must remain at the instruction site, and the deferral scan never crosses them.
%%%
\begin{algorithm}[t]
\caption{Post-Handler Deferral Algorithm}
\label{alg:post-handler-deferral}
\begin{algorithmic}[1]
\Require faulting instruction $I_f$, original fault metadata $F$, maximum scan distance \textsc{max\_scan}
\Ensure deferred post-handler address $\textit{safeAddr}$ and flag $\textit{postHandlerNeeded}$

\Statex
\Function{FindPostHandlerSafeSite}{$I_f, F, \textsc{max\_scan}$}
    \State $\textit{safeAddr} \gets \Call{NextInstruction}{I_f}$
    \State $\textit{postHandlerNeeded} \gets \mathsf{true}$
    \State $\textit{lastSimilarMemAddr} \gets \bot$
    \State $\textit{taggedRegs} \gets \Call{ExtractTaggedRegisters}{F}$
    \State $\textit{faultShape} \gets \Call{ExtractMemoryShape}{F}$
	\Comment memory access pattern of {$I_f$} (base/index/scale/displacement and access size)

    \For{$i \gets 1$ to \textsc{max\_scan}}
        \State $I \gets \Call{DecodeNextInstruction}{}$

        \If{\Call{IsControlFlowInstruction}{$I$}}
            \If{\Call{IsSingleExitLoopLatch}{$I, I_f$}}
                \State $\textit{safeAddr} \gets \Call{FallThroughOf}{I}$
                \Comment{conditional back-edge enclosing $I_f$: restore the tag once at the loop's single exit (gates in text)}
            \EndIf
            \State \Return $(\textit{safeAddr}, \textit{postHandlerNeeded})$
        \EndIf

        \If{\Call{IsTagIndependent}{$I, \textit{taggedRegs}$}}
            \If{\textbf{not} \Call{CanFaultOnHeap}{$I$} \textbf{and} $\textit{safeAddr} \leq $\textit{lastSimilarMemAddr}}
                \State $\textit{safeAddr} \gets \Call{AddressOf}{I}$
                \Comment{host site: register-only, \texttt{LEA}/\texttt{NOP}, or provably non-heap operands}
            \EndIf
            \State \textbf{continue}
        \EndIf

        \If{\Call{UsesTaggedMemoryAddress}{$I, \textit{taggedRegs}$}}
            \If{\textbf{not} \Call{SameMemoryPattern}{$I, \textit{faultShape}$}}
                \State $\textit{ea} \gets \Call{RecomputeEffectiveAddress}{$I$}$
				\If{\textbf{not} \Call{WithinOriginalValidBounds}{$\textit{ea}, F$}}
					\State \Call{ReportOOB}{$I, \textit{ea}, F$}
					\If{\Call{AbortOnOOB}{}}
						\State \Call{AbortExecution}{} \Comment{report OOB violation (unsafe divergence) and abort if in strict mode}
					\EndIf
					\State \Return $(\textit{safeAddr}, \textit{postHandlerNeeded})$
				\EndIf
            \EndIf
            \State $\textit{lastSimilarMemAddr} \gets \Call{AddressOf}{I}$
            \State \textbf{continue}
        \EndIf

        \If{\Call{FullyOverwrites}{$I, \textit{taggedRegs}$}}
            \State $\textit{postHandlerNeeded} \gets \mathsf{false}$
            \State \Return $(\textit{safeAddr}, \textit{postHandlerNeeded})$
        \EndIf

        \If{\Call{IsConstantPointerWalk}{$I, \textit{taggedRegs}$}}
            \State \Call{FoldIntoDelta}{$I$}
            \Comment{\texttt{add}/\texttt{sub} $r$, imm; \texttt{inc}/\texttt{dec} $r$: later fault-like accesses validate at the shifted address}
            \State \textbf{continue}
        \EndIf

        \State \Return $(\textit{safeAddr}, \textit{postHandlerNeeded})$
        \Comment{tag is consumed: tagged registers are read or partially overwritten}
    \EndFor

    \State \Return $(\textit{safeAddr}, \textit{postHandlerNeeded})$ \Comment{maximum scan distance reached; stop deferral conservatively}
\EndFunction
\end{algorithmic}
\end{algorithm}
%%%%%

\subsection{C Library and Kernel Boundaries}
Tagged pointers require special handling at libc and kernel boundaries. Kernel interfaces may reject noncanonical user addresses, while some libc routines perform pointer manipulations incompatible with the tagged representation. Hence, a tagged pointer dereferenced inside a system call or a glibc internal routine may trigger a fault that the signal handler cannot intercept, leaving the process in an unrecoverable state. MallocSan therefore wraps a broad set of libc entry points (memory, string, file, locale, and pthread routines), stripping tags from pointer arguments before each boundary crossing and reapplying them to any returned or derived pointer that should remain associated with the same protected object. This wrapper layer is a structural part of the design, not an optional convenience: many library operations perform internal pointer arithmetic or forward user buffers to the kernel, making transparent tag removal mandatory. Scoped coverage is an additional benefit: by selectively wrapping or excluding specific entry points, MallocSan confines protection to application components while leaving standard shared libraries uninstrumented.

This design is complemented by a reentrancy guard that prevents MallocSan's internal allocations, as well as those issued within wrapped libraries, from being subjected to tagging. The guard is implemented as a thread-local Boolean flag. Upon entry to each interposed allocator or \texttt{libc} wrapper, the flag is saved and cleared, temporarily disabling protection for internally issued allocations. On exit, the flag is restored to its prior state. Without this guard, an internal allocation would re-enter the interposed allocator with protection enabled, tag its own result, fault on the tagged pointer, and recurse until exhausting either the call stack or the OID table. Declaring the flag thread-local ensures that suppressing instrumentation in one thread does not inadvertently disable protection in any other.

\subsection{Concurrency and Thread Safety}

MallocSan achieves thread safety by separating immutable instruction metadata from mutable per-execution state. The instruction table stores per-instruction properties (decoded operands, register-access summaries, patching mode, and post-handler placement) that are written once and read frequently. Mutable state resides in a thread-local runtime structure holding temporary register values during execution. This separation eliminates contention on the instruction table along the common read path and prevents transient execution data from corrupting persistent metadata.

Concurrency requires coordination at two points. First, instruction-table entries are created through an atomic state machine that ensures exactly one thread decodes a newly faulting instruction while all others encountering the same site wait for the entry to reach the ready state. Second, once the process becomes multithreaded, patch requests are serialized through a ring buffer to a dedicated patch worker, satisfying \texttt{libpatch}'s single-threaded patching constraint without blocking application threads during ordinary execution. Performing instruction analysis outside the signal handler is also necessary for thread safety: because Capstone is not async-signal-safe, each thread maintains an independent disassembly context, and executing the analysis outside the handler guarantees that these contexts are never accessed from within a signal.

\subsection{Guard-Page Protection Backend}\label{subsec:guard-backend}
\begin{figure*}
\centerline{
\includegraphics[width=0.75\textwidth]{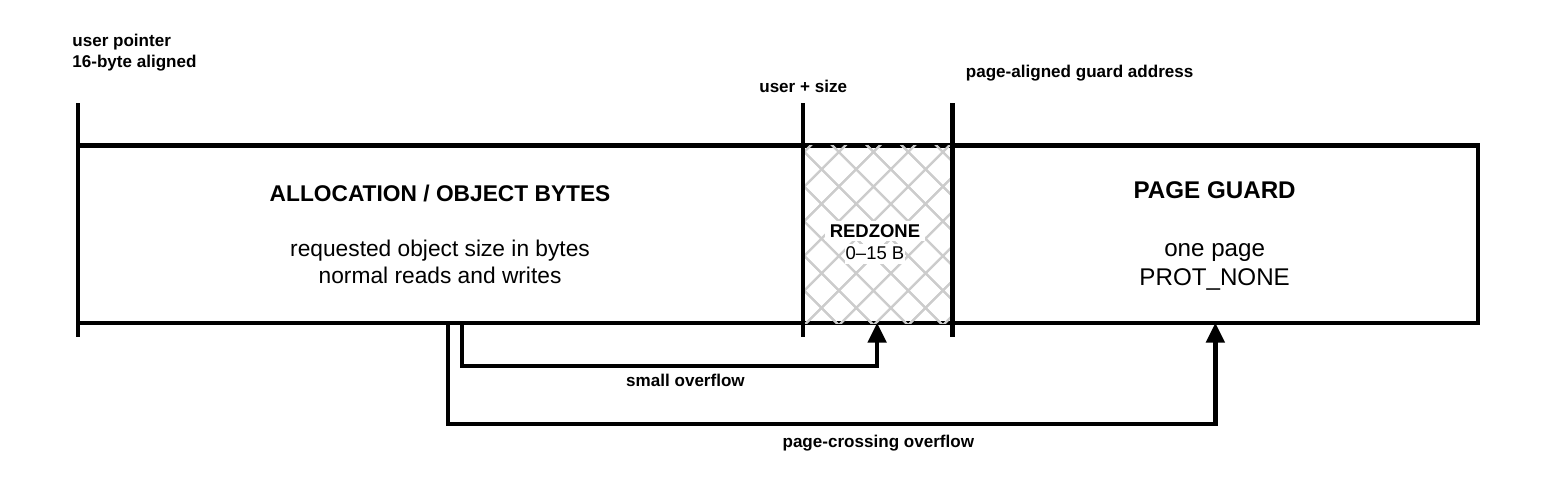}}
\caption{Memory layout of a guard-page allocation. The object is placed at the end of its mapping, immediately before a \texttt{PROT\_NONE} page, so a page-crossing overflow faults at the offending access, while the 16-byte alignment gap is filled with canary bytes that are verified at deallocation to catch small trailing overflows. \label{fig:guard-backend}}
\Description{Diagram of a guard-page allocation as three adjacent memory regions. A block of object bytes starts at the 16-byte-aligned user pointer, followed at the address user plus size by a redzone of zero to fifteen canary bytes, followed at a page-aligned address by one inaccessible PROT_NONE guard page. Two arrows leave the object block: a small overflow lands in the redzone, and a page-crossing overflow lands on the guard page.}
\end{figure*}

In addition to OID tagging, MallocSan offers an optional \emph{guard-page} backend that protects selected allocations without any per-access instrumentation. As shown in Fig.~\ref{fig:guard-backend}, an eligible allocation is served from a dedicated anonymous \texttt{mmap} region and placed at its end, so that the object is immediately followed by an inaccessible \texttt{PROT\_NONE} page, in the manner of Electric Fence~\cite{perens1993efence}. Any access that runs past the object lands on the guard page and raises a hardware fault at the exact offending instruction. Since the returned pointer is a plain canonical pointer rather than a tagged one, dereferences of guarded objects never fault during ordinary execution and bypass the fault-handling, analysis, and patching machinery of the preceding subsections entirely.

This model gives up a small amount of precision. Preserving the 16-byte alignment that \texttt{malloc} guarantees can leave a gap of up to 15 bytes between the object's last byte and the guard page. Overflows that reach the page, including every overflow of 16 bytes or more, trap immediately. To mitigate this limitation, the backend fills the residual gap with canary bytes at allocation and verifies them when the object is freed or resized in place, thereby detecting trailing writes that modify the gap. A small trailing read within the gap goes undetected, and underflows are outside this backend's reach, since only the end of the allocation is guarded. The OID backend, in contrast, detects all of these cases at the moment of access.

The memory cost is one page-granular mapping plus one inaccessible page per guarded object, so guarding is gated by a configurable minimum allocation size and a cap on concurrently live guarded objects. The two backends operate side by side: requests at or above the size threshold go to the guard backend, and all others, as well as any request it declines, fall back transparently to OID tagging. Guarded pointers are recognized ahead of the tagged-pointer logic on both the \texttt{free} path and the fault path, and a guard fault produces the same style of report as the OID path: object bounds, allocation site, and overflow distance. This division of labor is deliberate. Large objects, for which page rounding is proportionally negligible, are precisely the ones whose dense access streams are most expensive to check through tagging, while small, numerous allocations, for which a page per object would be wasteful, remain under the OID model. Shifting large objects to hardware-enforced guard pages thus eliminates much of the per-access overhead at a minor cost in accuracy and memory.

\section{Implementation Details}\label{sec:implementation}
The MallocSan implementation is organized around four runtime modules that mirror the main phases of execution. The \emph{runtime-control} module initializes the sanitizer, parses the environment-controlled protection policy, and installs the signal handlers. The \emph{instruction-analysis} module decodes faulting instructions, caches instruction-table entries, and implements the pre-handler and post-handler logic of the steady-state path, including the forward scan for post-handler elimination and deferral. The \emph{patching} module interfaces with \texttt{libpatch} to install binary rewrites, and routes patch requests to the patch worker. Finally, the \emph{wrapper} module interposes the \texttt{malloc} family and selected libc entry points, tagging newly protected allocations and stripping or restoring tags at library and kernel boundaries.

The implementation comprises more than 10{,}000 lines of C code and is built with GNU Autotools using GCC or Clang on x86-64 Linux. It depends on four external runtime libraries: \texttt{libpatch} for binary patching, Capstone for disassembly, and \texttt{libdw}/\texttt{libunwind} for ELF and DWARF introspection. Beyond detecting violations, MallocSan is designed to explain them. Object-table entries retain the allocation-site instruction pointer, so an out-of-bounds report identifies both the violated object and the offending access. The runtime also records per-instruction hit counts and patching metadata, and emits a statistics file at process exit listing each instruction-table entry with its address, symbolized location, access count, and patching mode. This file drives the profile-guided optimization of Section~\ref{sec:pgo}.

\section{Profile-Guided Hot-Path Optimization}\label{sec:pgo}

Although \texttt{TRAP} patching provides a safe and nearly universal fallback when a direct redirection cannot be installed at runtime, it is an expensive steady-state execution mode: every invocation incurs signal delivery, handler dispatch, and context recovery. In-handler emulation absorbs the simplest of these sites (Section~\ref{subsec:patching-fallbacks}), but instructions too complex to emulate remain trap-bound, and their cost is particularly harmful inside hot loops, where repeated traps can dominate total execution time. MallocSan therefore includes an optional profile-guided optimization pipeline that identifies frequently executed \texttt{TRAP} sites and rewrites them offline into direct \texttt{JUMP}-based redirections. The goal is to preserve checking semantics while promoting hot sites from a fault-driven slow path to a normal user-space control-flow path.

The optimization relies on E9Patch~\cite{duck2020binary}, a static binary rewriter for optimized \mbox{x86-64} ELF executables, and proceeds in two phases. In the \emph{profiling phase}, the target program runs under MallocSan with its default runtime patching logic and produces the per-instruction statistics file described in Section~\ref{sec:implementation}. Its counters directly reveal which sites remain on the \texttt{TRAP} path and how frequently each executes.

In the \emph{rewriting phase}, a Python script reads the statistics file and selects candidate instructions according to three criteria: the site must have been finalized as a \texttt{TRAP}-patched instruction during profiling, its execution count must exceed a user-specified threshold, and the instruction must offer a safe rewrite window, that is, a contiguous span of at least five bytes whose relocation captures no additional memory-access instruction. For the selected subset, the script invokes E9Patch to produce a new binary in which each eligible \texttt{TRAP} site is replaced by a \texttt{JUMP}-based redirection performing the same OID stripping, bounds check, and tag restoration as the original signal-driven path. The rewritten binary still runs with MallocSan preloaded, since the injected handlers continue to consult the object table and other runtime metadata maintained by the sanitizer.

This selective rewriting preserves correctness because the rewritten fast path reuses the same checking logic as the original \texttt{TRAP}-based path; only the dispatch mechanism changes. Instructions that do not satisfy the eligibility criteria remain unchanged and continue to use the signal-based fallback. The approach therefore trades the cost of one profiling run for lower steady-state overhead on hot sites in subsequent executions, making it particularly attractive for long-running workloads and benchmark settings in which hot-path behavior is stable across runs.

\section{Evaluation}\label{sec:evaluation}

We organize the evaluation around four research questions:

\begin{description}
  \item[RQ1: Detection effectiveness.] Does MallocSan report an out-of-bounds violation for every heap allocation-boundary error that manifests within its protection scope, and does it remain silent on the corresponding defect-free executions?
  \item[RQ2: Runtime cost.] What execution-time and memory overhead does MallocSan impose on SPEC CPU~2017 workloads, relative to Native execution and Memcheck?
  \item[RQ3: Profile-guided rewriting.] How much does offline rewriting reduce the cost of workloads dominated by hot \texttt{TRAP}-based sites?
  \item[RQ4: Multithreaded scaling.] How does MallocSan's overhead vary with thread count, and does it degrade the application's parallel speedup relative to Native?
\end{description}

For the performance evaluation, we compare MallocSan with Valgrind's Memcheck~\cite{nethercote2007valgrind}, a binary-level memory-error detector that can analyze unmodified, potentially closed-source x86-64 binaries without recompilation or specialized hardware. ASan is widely used and provides an important practical point of comparison, but its compile-time instrumentation normally requires rebuilding the application~\cite{serebryany2012addresssanitizer}. Memcheck is therefore the more directly comparable performance baseline in terms of deployment model, although it checks memory addressability broadly while MallocSan focuses on accesses through protected heap pointers. To reduce work unrelated to spatial-access validation, we disabled Memcheck's leak and undefined-value error reporting and used the remaining configuration shown below.\footnote{Memcheck was invoked with \texttt{-{}-leak-check=no}, \texttt{-{}-undef-value-errors=no}, \texttt{-{}-track-origins=no}, and \texttt{-{}-partial-loads-ok=yes}.}

\subsection{Detection Effectiveness (RQ1)}\label{sec:eval-accuracy}
We assessed detection accuracy using version~1.3 of the NIST Juliet Test Suite for C/C++~\cite{black2018juliet}. The analysis covers the allocator-backed bounds-error families of CWE-122 (heap-based buffer overflow), CWE-124 (buffer underwrite), CWE-126 (buffer overread), and CWE-127 (buffer under-read). We included a test-case family only when its source obtains the affected buffer through \texttt{malloc}, \texttt{calloc}, \texttt{realloc}, \texttt{memalign}, or \texttt{posix\_memalign}, consistent with MallocSan's heap-oriented protection scope.

This selection produced 2{,}766 test-case families. Each family contributed two executions: a \emph{good} run, which exercises the test case without the seeded defect, and a \emph{bad} run, which attempts to trigger the defect. The initial corpus therefore contained 5{,}532 executions. Each binary ran with MallocSan preloaded and with deterministic inputs, including fixed out-of-range indices and controlled file, environment, random, and socket fixtures. The oracle classified a manifested, in-scope bad run as detected only when MallocSan emitted its out-of-bounds diagnostic; any such diagnostic during a good run counted as a false positive.

Before scoring RQ1, we reviewed the 571 bad runs that emitted no MallocSan
diagnostic. For each run, we established ground truth by manually inspecting
the executed Juliet path under the deterministic input and evaluating the
target object and accessed byte range under the x86-64 ABI. The missing
diagnostic triggered the review, but exclusion required evidence that no
violation manifested or that the access did not cross the boundary of an
eligible heap allocation.
This review identified 152 runs in which the seeded defect did not manifest:
144 \texttt{sizeof}-confusion cases were inactive on the 64-bit ABI because the
erroneous pointer-size expression equaled the intended pointee size, and eight
runs deterministically selected a safe, non-vulnerable sink. The remaining 419
runs contained violations outside the scope of heap-allocation bounds checking:
383 overflowed a stack-resident destination buffer even though the source
buffer was allocated on the heap, and 36 crossed a field boundary while
remaining within the enclosing heap allocation. The latter are intra-object
violations that an allocation-granularity detector cannot distinguish from
valid accesses without subobject metadata. We retain all 2{,}766 good runs as
negative cases.

\begin{table}[t]
  \centering
  \caption{RQ1 outcomes on the scored population. A positive diagnostic is MallocSan's out-of-bounds report.}
  \label{tab:juliet-confusion}
  \begin{tabular}{lrrr}
    \toprule
    Run class & Report & No report & Total \\
    \midrule
    In-scope bad & 2{,}195 & 0 & 2{,}195 \\
    Good         & 0       & 2{,}766 & 2{,}766 \\
    \midrule
    Total        & 2{,}195 & 2{,}766 & 4{,}961 \\
    \bottomrule
  \end{tabular}
\end{table}

Table~\ref{tab:juliet-confusion} gives the resulting confusion matrix. MallocSan produced 2{,}195 true positives and 2{,}766 true negatives, with no false negatives or false positives. Accuracy, precision, recall, specificity, and the $F_1$ score are therefore all 1.00 on this scored population. The population includes reads and writes as well as underflows and overflows; the OID model treats these cases symmetrically by validating every protected access against both bounds of its referenced allocation.

Beyond the Juliet suite, we evaluated MallocSan on existing applications and detected multiple heap-buffer underflows and overflows. For example, MallocSan reproduced the one-byte heap-buffer overread in \emph{LibTIFF 4.4.0} reported in issue \#435 and CVE-2022-3598 \cite{libtiff_issue435,nvd_cve20223598}. When \emph{tiffcrop} processed the malformed TIFF file with \emph{-Z 1:4,3:3}, MallocSan identified the invalid read in \emph{extractContigSamplesShifted32bits()} immediately beyond a 120-byte allocation created by \emph{rotateImage()}. ASan independently reported the same invalid-access and allocation call stacks, confirming the result. The violation was no longer detected with the patched \emph{LibTIFF 4.5.0}. This case shows that MallocSan can detect byte-level spatial errors in real applications and provide diagnostic information consistent with ASan.

\paragraph{Answer to RQ1.}
For every manifested allocation-boundary violation in the selected Juliet population, MallocSan emitted its out-of-bounds diagnostic, and it emitted no such diagnostic in any corresponding good execution. This result is conditional on the stated oracle and scope: it does not cover the 152 non-manifesting bad runs, the 383 stack-destination violations, or the 36 intra-object violations. Our evaluation on existing applications provides complementary evidence: MallocSan detected multiple heap-buffer underflows and overflows. In the \emph{LibTIFF case}, it detected the same one-byte overread and reported the same access and allocation call stacks as ASan, indicating that its detection capability extends to real software.

\subsection{Performance Methodology}\label{sec:eval-performance}

All experiments were conducted on a machine equipped with a 12th~Gen Intel Core
i7-12700 processor (8 performance cores, 4 efficiency cores, and 20 hardware
threads; 2.1~GHz performance-core base frequency; 25~MB L3 cache) and 16~GB of
RAM. The machine ran Ubuntu~22.04 with Linux kernel~6.8 and glibc~2.35, and
processor Turbo Boost was disabled throughout the experiments. We compiled the
benchmarks with GCC~11.4.0 using \texttt{-O3} together with the required SPEC
portability flags. We used Valgrind~3.21.0 with the Memcheck options listed
earlier. We ran the benchmarks with the SPEC reference (\texttt{refspeed}) input
set, invoking the binaries directly with the reference command lines, and we
verified outputs with SPEC's \texttt{specdiff} comparison tool.

MallocSan's allocation-selection policy was configured to protect every eligible heap allocation using the OID-based backend described in Section~\ref{sec:design}. The experiments used default OID recycling and did not enable optional use-after-free tracking, whose detection effectiveness and overhead are outside the scope of this evaluation. We also disabled the guard-page backend described in Section~\ref{subsec:guard-backend} because it reduces per-access checking costs at the expense of some detection precision. Evaluating only the OID-based backend reports MallocSan's overhead under its most precise configuration and provides a more conservative comparison with Memcheck's per-access addressability checking.

% PIGZ UPDATE: This paragraph now scopes the seven benchmarks to the SPEC evaluation.
The single-thread SPEC performance evaluation covers seven benchmarks from the SPEC CPU~2017 speed suites: 605.mcf\_s, 619.lbm\_s, 620.omnetpp\_s, 625.x264\_s, 631.deepsjeng\_s, 644.nab\_s, and 648.exchange2\_s. Respectively, these workloads represent minimum-cost-flow computation with intensive pointer and integer arithmetic, lattice-Boltzmann fluid simulation, discrete-event network simulation, video encoding, chess-tree search, molecular modeling, and recursive Sudoku generation. Benchmark selection required that each workload both exercise heap accesses covered by MallocSan's allocator-interposition model and execute correctly with its high-bit pointer representation. Accordingly, these results characterize MallocSan on selected workloads that yield application-level heap-allocation candidates under its current policy, rather than on the entire SPEC CPU 2017 suite. For example, we excluded 998.specrand\_s because it made no application-level heap allocations. The only observed allocation was a 4 KiB shared-library allocation. Other benchmarks were incompatible with high-bit tagging. For example, 602.gcc\_s uses GCC's garbage-collected allocator, whose internal page-table lookup interprets the raw pointer bit pattern and consequently misinterprets tagged pointers, as discussed in Section~\ref{sec:limitations}.

%For 641.leela\_s, MallocSan reported zero protected allocations and zero candidates, but this does not indicate that the benchmark avoids heap memory. An independent heap profile recorded 601,747 malloc calls, allocating about 1.47 GB cumulatively with a 21.2 MB peak heap. The benchmark is written in C++ and uses new, std::vector, smart pointers, and related facilities. These requests reach malloc through libstdc++; MallocSan’s current caller-address filter excludes allocations whose immediate caller is in a shared library, so they are not considered protection candidates.

\begin{figure}[t]
  \centering
  \includegraphics[width=0.6\linewidth]{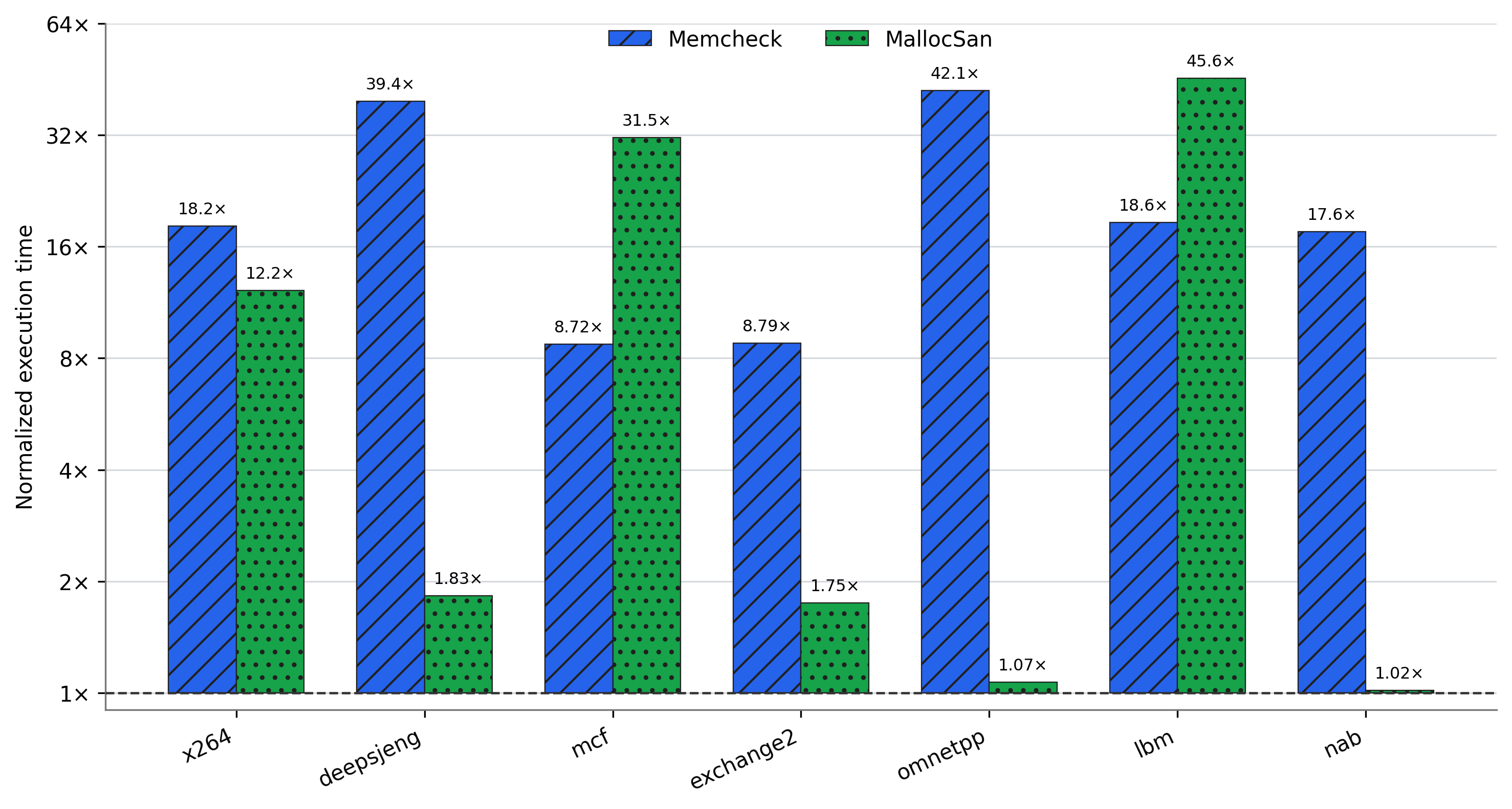}
  \caption{Execution-time overhead of Valgrind Memcheck and MallocSan on seven SPEC CPU~2017 benchmarks. Each factor is the tool's mean execution time divided by the corresponding Native mean; the vertical axis uses a base-2 logarithmic scale. The dashed line marks the $1.0\times$ Native baseline, and lower is better.}
  \Description{Grouped bars compare Valgrind and MallocSan execution time normalized to Native for seven benchmarks. MallocSan is lower on x264, deepsjeng, exchange2, omnetpp, and nab, while Valgrind is lower on mcf and lbm.}
  \label{fig:exec-overhead}
\end{figure}

\begin{figure}[t]
  \centering
  \includegraphics[width=0.6\linewidth]{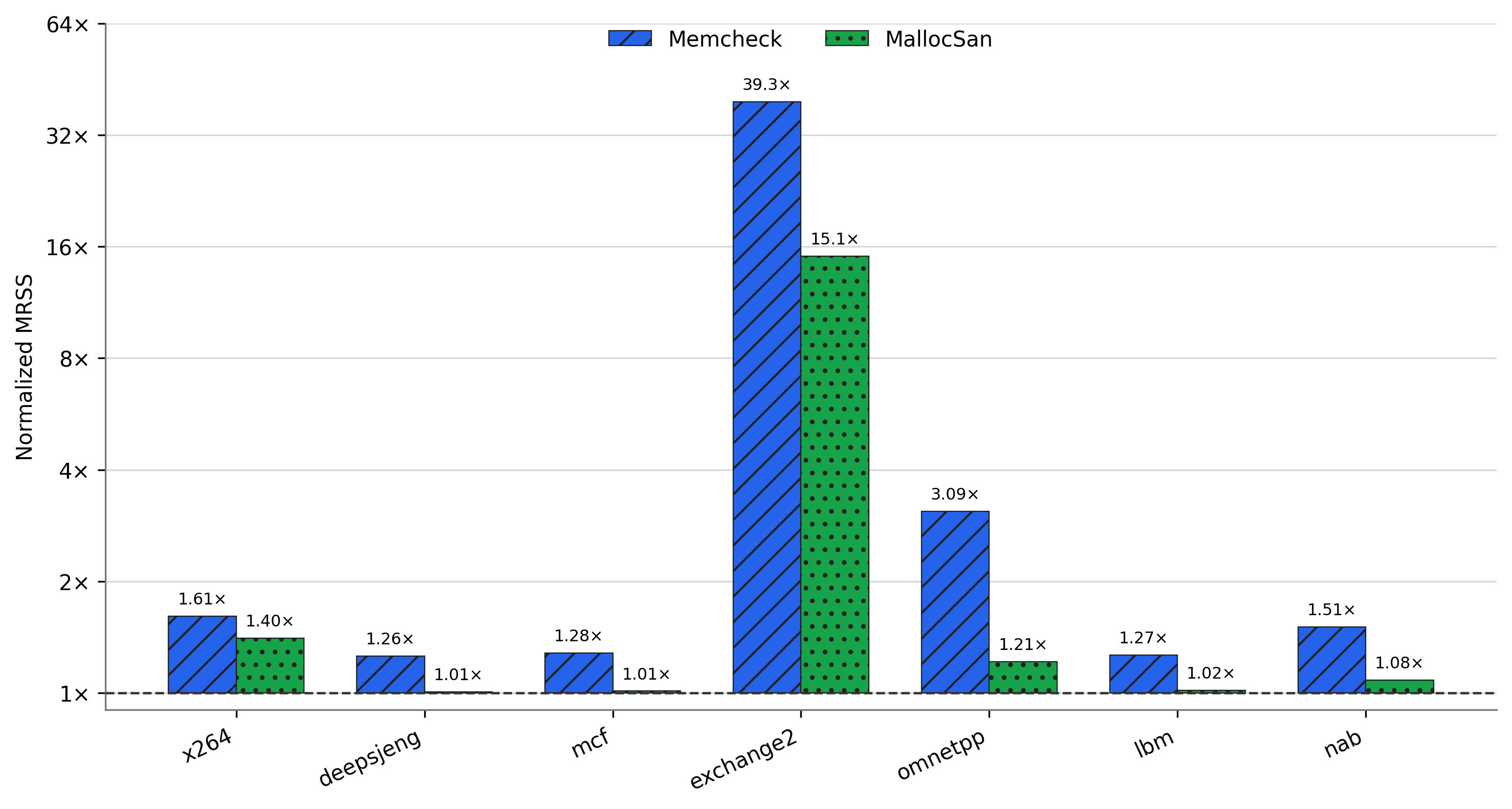}
  \caption{Normalized MRSS of Memcheck and MallocSan on seven selected SPEC CPU~2017 benchmarks. Each factor is the tool's mean MRSS divided by the corresponding Native mean. The vertical axis uses a base-2 logarithmic scale; the dashed line marks the $1.0\times$ Native baseline, and lower is better.}
  \Description{Grouped bars compare the maximum resident set sizes of Memcheck and MallocSan, normalized to Native, on seven benchmarks. MallocSan uses less resident memory on every benchmark; exchange2 has the largest normalized memory factor for both tools.}
  \label{fig:mem-overhead}
\end{figure}

All measurements reported in Figures~\ref{fig:exec-overhead} and~\ref{fig:mem-overhead} used one application thread; the separate scaling experiments in Section~\ref{sec:eval-scaling} varied the thread count for 644.nab\_s and pigz. For this single-thread SPEC experiment, we executed each benchmark configuration ten times. For each run, \texttt{/usr/bin/time} recorded wall-clock execution time and maximum resident set size (MRSS). All Native, Memcheck, and MallocSan executions passed SPEC's output validation. We report the arithmetic mean of each quantity across the ten runs.

\[
R_{b,u,m}
=
\frac{\overline{x}_{b,u,m}}
     {\overline{x}_{b,\mathrm{Native},m}},
\qquad
u \in \{\mathrm{Memcheck},\mathrm{MallocSan}\}.
\]

Here, $R_{b,u,m}$ is the normalized factor and $\overline{x}_{b,u,m}$ is the arithmetic mean for benchmark $b$, tool $u$, and measured quantity $m$, with $m$ denoting either execution time or MRSS. Because the denominator is the mean for the uninstrumented Native configuration, Native defines a $1.0\times$ baseline, and larger factors indicate greater overhead. Aggregate factors are computed as the geometric mean of the per-benchmark normalized factors. Figures~\ref{fig:exec-overhead} and~\ref{fig:mem-overhead} use base-2 logarithmic vertical axes so that both modest and extreme factors remain visible within each plot. Each bar is annotated with its normalized value, and the dashed horizontal line marks the Native baseline.

Section~\ref{sec:discussion-findings} interprets these results and examines the design tradeoffs they expose.

\subsection{Single-Thread Overhead (RQ2)}\label{sec:eval-single-thread}

\paragraph{Execution time.}
Figure~\ref{fig:exec-overhead} reports execution time normalized to Native. MallocSan has a lower factor than Memcheck on five of the seven benchmarks. Its overhead is close to Native on 644.nab\_s ($1.02\times$) and 620.omnetpp\_s ($1.07\times$), and remains below $2\times$ on 648.exchange2\_s ($1.75\times$) and 631.deepsjeng\_s ($1.83\times$). On these four workloads, Memcheck's factors range from $8.79\times$ to $42.14\times$. MallocSan also outperforms Memcheck on 625.x264\_s, with factors of $12.20\times$ and $18.19\times$, respectively.

Memcheck has lower factors on the other two benchmarks. On 605.mcf\_s, MallocSan reaches $31.46\times$ Native execution time, compared with $8.72\times$ for Memcheck. We performed a static disassembly of the evaluated 605.mcf\_s executable to characterize its control-flow density. The disassembly identifies 1{,}127 control-flow instructions among 6{,}463 instructions ($17.4\%$). On 619.lbm\_s, MallocSan reaches $45.56\times$ Native execution time, compared with $18.57\times$ for Memcheck.

Across the seven selected benchmarks, MallocSan's geometric-mean factor is $4.82\times$, compared with $18.55\times$ for Memcheck. Memcheck's geometric-mean factor is therefore approximately $3.85\times$ as large as MallocSan's.

\paragraph{Memory.}
Figure~\ref{fig:mem-overhead} reports the normalized MRSS factors. MallocSan has a lower mean MRSS than Memcheck on every evaluated benchmark. On six workloads, its factors range from $1.01\times$ to $1.40\times$: $1.01\times$ on 631.deepsjeng\_s and 605.mcf\_s, $1.02\times$ on 619.lbm\_s, $1.08\times$ on 644.nab\_s, $1.21\times$ on 620.omnetpp\_s, and $1.40\times$ on 625.x264\_s. The corresponding Memcheck factors range from $1.26\times$ to $3.09\times$. Across all seven benchmarks, the geometric-mean MRSS factor is $1.62\times$ for MallocSan and $2.50\times$ for Memcheck.

The outlier in normalized terms is 648.exchange2\_s, with factors of $15.09\times$ for MallocSan and $39.33\times$ for Memcheck. The benchmark's Native footprint is 3.59~MB. The mean MRSS values are 54.18~MB for MallocSan and 141.20~MB for Memcheck, corresponding to absolute increases over Native of approximately 50.6~MB and 137.6~MB, respectively. As a sensitivity analysis, excluding this small-denominator workload yields geometric-mean factors of $1.11\times$ for MallocSan and $1.58\times$ for Memcheck across the remaining six benchmarks; the seven-benchmark geometric means remain the primary aggregate results.

\paragraph{Answer to RQ2.}
On the seven selected compatible workloads, MallocSan has a lower mean execution-time factor than Memcheck on five benchmarks and a lower memory overhead on all seven. Its geometric-mean execution-time overhead is $4.82\times$ Native, compared with $18.55\times$ for Memcheck. Its geometric-mean MRSS factor is $1.62\times$ Native, compared with $2.50\times$ for Memcheck. These aggregate results characterize the selected workloads under the broadest OID policy and should not be interpreted as estimates for the complete SPEC CPU~2017 suite.

\subsection{Effect of Profile-Guided Rewriting (RQ3)}\label{sec:eval-pgo}
Because 605.mcf\_s and 619.lbm\_s exhibit the two highest MallocSan slowdowns, we use them as a targeted case study of the profile-guided pipeline described in Section~\ref{sec:pgo}. We first ran each benchmark under MallocSan to collect per-site execution counts and identify frequently executed instructions that remained on the \texttt{TRAP} path. The offline rewriting phase then used E9Patch~\cite{duck2020binary} to replace each eligible hot site with a direct \texttt{JUMP}-based redirection. A site was eligible only if it provided a safe rewrite window spanning at least five contiguous bytes and relocating that window captured no additional memory-access instruction. The rewritten binaries continued to run with MallocSan preloaded and performed the same OID stripping, bounds validation, and tag restoration as the original \texttt{TRAP}-based path; only the dispatch mechanism changed.

Profile-guided rewriting reduced the normalized execution time of both workloads, although by different amounts. For 605.mcf\_s, the factor decreased from $31.46\times$ to $26.50\times$, a $15.8\%$ reduction. For 619.lbm\_s, it decreased from $45.56\times$ to $23.40\times$, a $48.6\%$ reduction. Across the two benchmarks, the geometric-mean factor decreased from $37.86\times$ to $24.90\times$, a $34.2\%$ reduction. For both workloads, the rewritten factor remained above the corresponding Memcheck factor. Both rewritten binaries passed SPEC's output validation for the evaluated inputs.

Because this targeted case study profiles and evaluates the same repeatable benchmark workloads, it estimates the attainable steady-state benefit when hot-path behavior is stable. It does not establish that the selected rewrite sites or speedups generalize to unseen inputs.

\paragraph{Answer to RQ3.}
Offline rewriting reduces normalized execution time by $15.8\%$ on 605.mcf\_s and $48.6\%$ on 619.lbm\_s, for a two-workload geometric-mean reduction of $34.2\%$. It mitigates, but does not eliminate, the overhead of these two unfavorable workloads; both rewritten factors remain above Memcheck's corresponding factors.

\subsection{Multithreaded Scaling (RQ4)}\label{sec:eval-scaling}
To evaluate both detector overhead and application scaling, we used the
OpenMP-enabled SPEC CPU~2017 benchmark 644.nab\_s and
pigz~\cite{adler2023pigz}, a parallel implementation of gzip that compresses
input blocks using multiple worker threads. Pigz complements the molecular
modeling benchmark with a general-purpose compression application used outside
benchmark suites. Of the two SPEC workloads with the highest MallocSan
overhead, 605.mcf\_s is single-threaded, while 619.lbm\_s scales poorly even
without instrumentation: its Native execution time falls from 1947.5~s with
one thread to 1178.1~s with 18 threads, a $1.65\times$ speedup. These workloads
therefore provide limited evidence about preserving substantial application
parallelism.

\begin{figure*}[t]
  \centering
  \begin{minipage}[t]{0.49\textwidth}
    \centering
    \includegraphics[width=\linewidth]{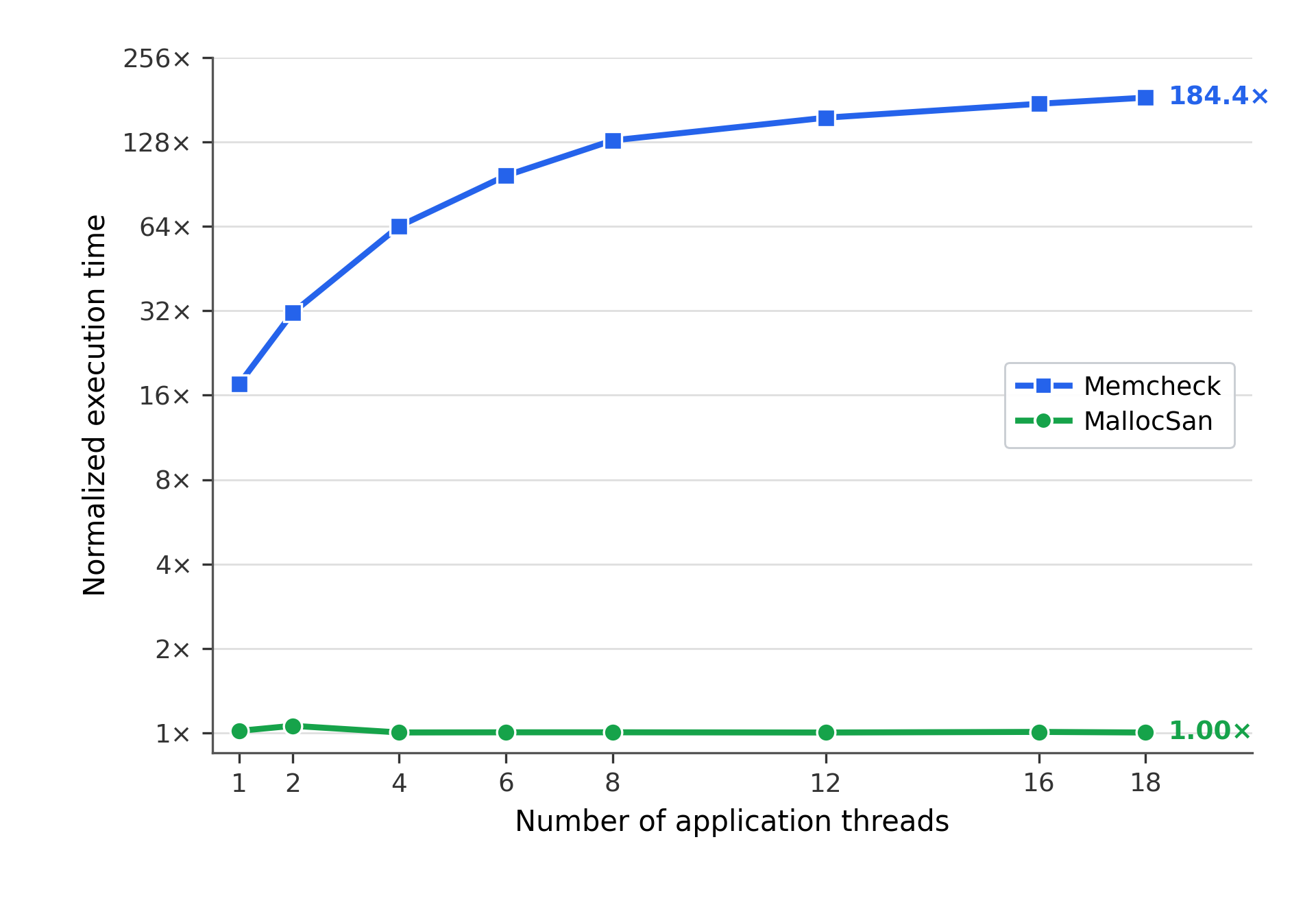}
    \par\smallskip\textbf{(a)} 644.nab\_s
  \end{minipage}\hfill
  \begin{minipage}[t]{0.49\textwidth}
    \centering
    \includegraphics[width=\linewidth]{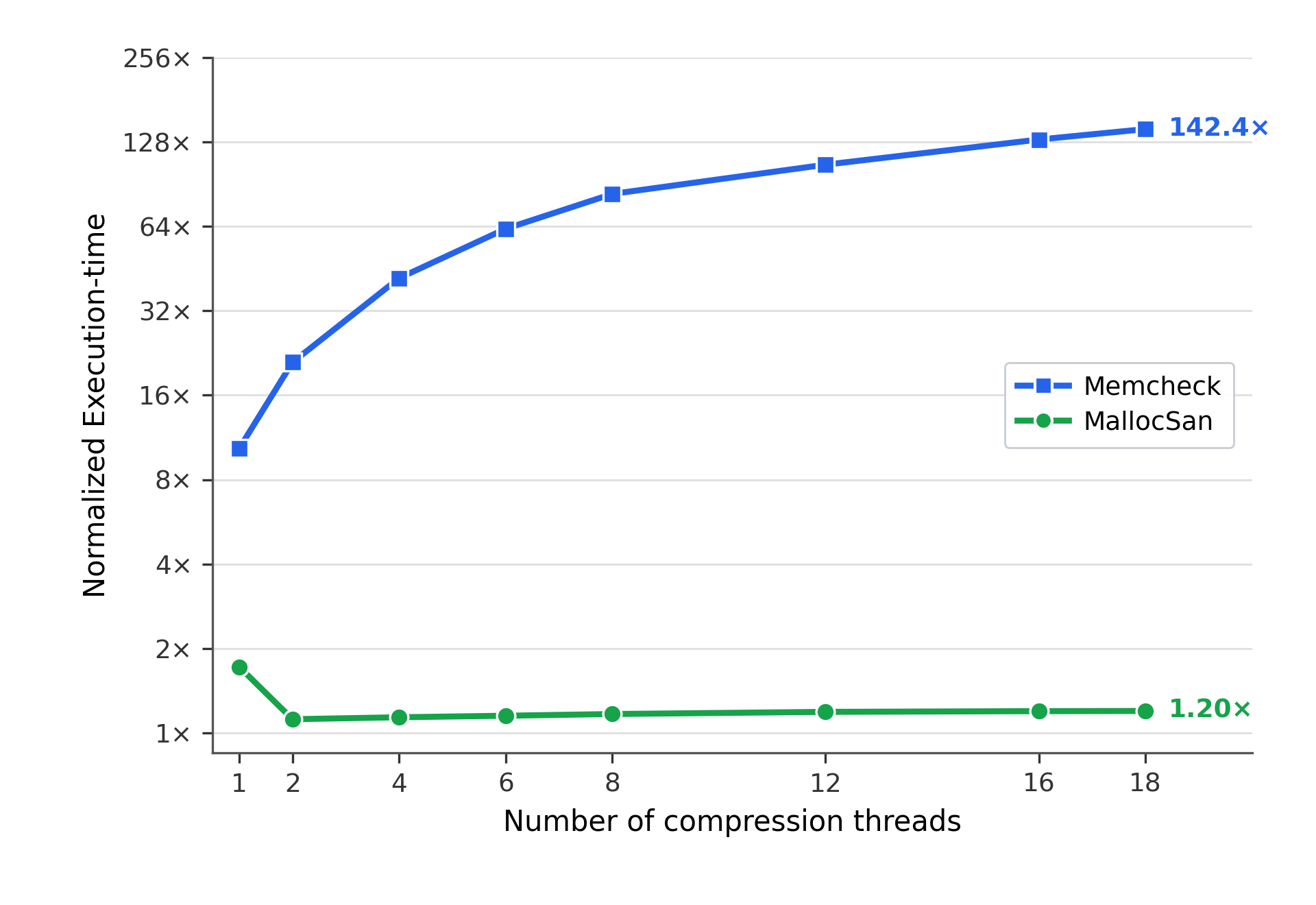}
    \par\smallskip\textbf{(b)} pigz
  \end{minipage}
  \caption{Execution-time overhead of Memcheck and MallocSan, normalized at each thread count to the corresponding Native execution time on a base-2 logarithmic scale; lower is better. Thread counts denote (a) OpenMP application threads for 644.nab\_s and (b) compression threads for pigz.}
  \Description{Two line charts show execution time normalized to Native across OpenMP thread counts for 644.nab_s and compression-thread counts for pigz. MallocSan remains near Native on nab. On pigz, its factor falls from 1.72 at one compression thread to 1.12 at two, then rises gradually to 1.20 at eighteen. Memcheck's normalized overhead increases substantially with the thread count on both workloads.}
  \label{fig:threads-overhead}
\end{figure*}

For both workloads, we swept $T\in\{1,2,4,6,8,12,16,18\}$ under Native,
Memcheck, and MallocSan. Here, $T$ denotes OpenMP application threads for
644.nab\_s and compression threads for pigz, not pigz's total process thread
count, which can also include I/O and detector helper threads. The maximum
configured count of 18 is below the processor's 20 hardware contexts, but is
not a cap on the total number of software threads. We invoked pigz as
\texttt{pigz -6 -p T -c input.tar}, keeping the compression level fixed at~6
and using the same approximately 1.00~GB input archive at every setting. The
\texttt{-p} option selects the compression-thread count; \texttt{-p 1} uses
pigz's non-threaded compression path~\cite{adler2023pigz}. Each pigz
configuration was run five times, and we report arithmetic-mean wall-clock
execution times. Figure~\ref{fig:threads-overhead} normalizes each detector's
mean against the Native mean for the same workload and thread count.

\begin{figure*}[t]
  \centering
  \begin{minipage}[t]{0.49\textwidth}
    \centering
    \includegraphics[width=\linewidth]{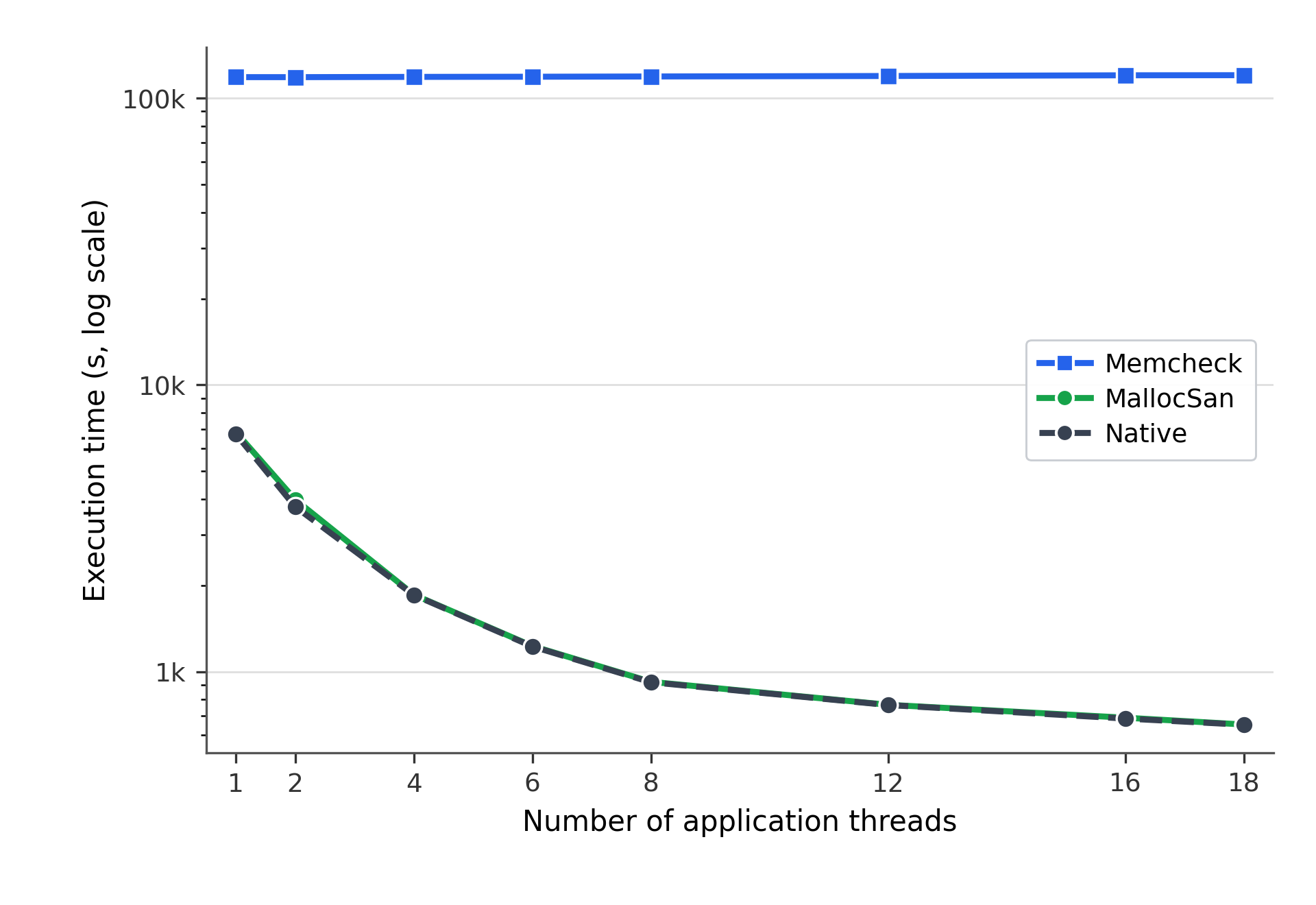}
    \par\smallskip\textbf{(a)} 644.nab\_s: execution time
  \end{minipage}\hfill
  \begin{minipage}[t]{0.49\textwidth}
    \centering
    \includegraphics[width=\linewidth]{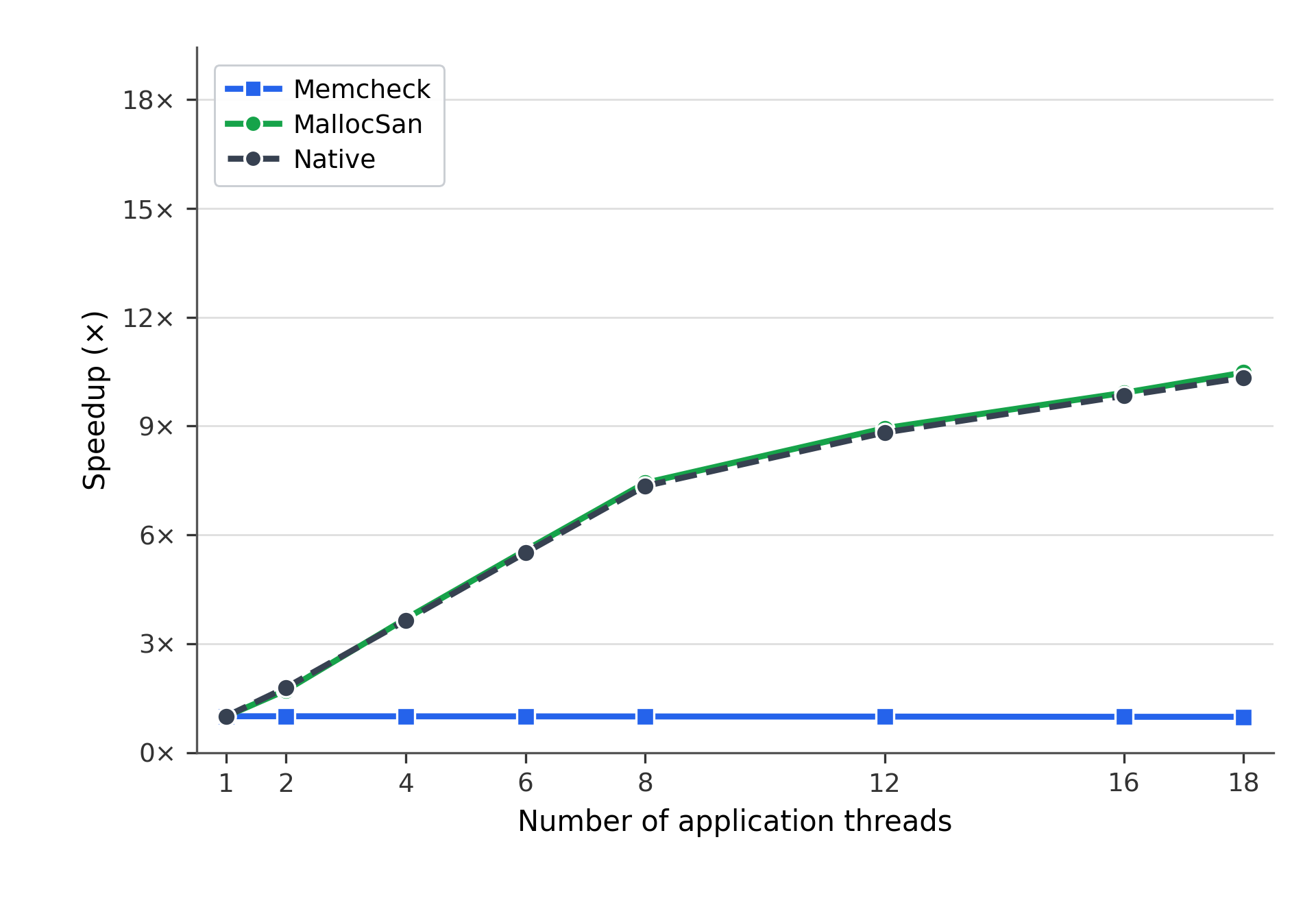}
    \par\smallskip\textbf{(b)} 644.nab\_s: speedup
  \end{minipage}
  \par\medskip
  \begin{minipage}[t]{0.49\textwidth}
    \centering
    \includegraphics[width=\linewidth]{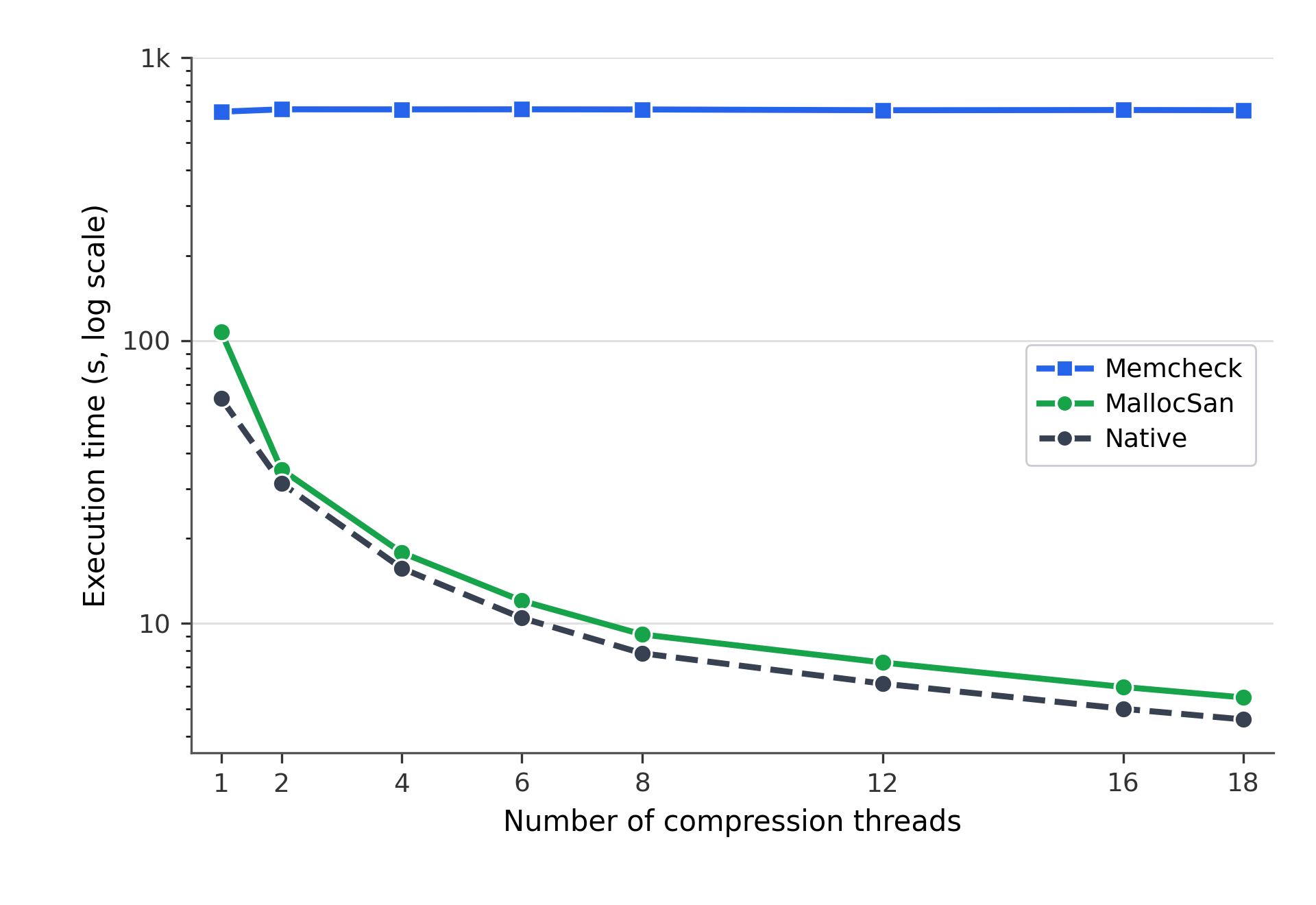}
    \par\smallskip\textbf{(c)} pigz: execution time
  \end{minipage}\hfill
  \begin{minipage}[t]{0.49\textwidth}
    \centering
    \includegraphics[width=\linewidth]{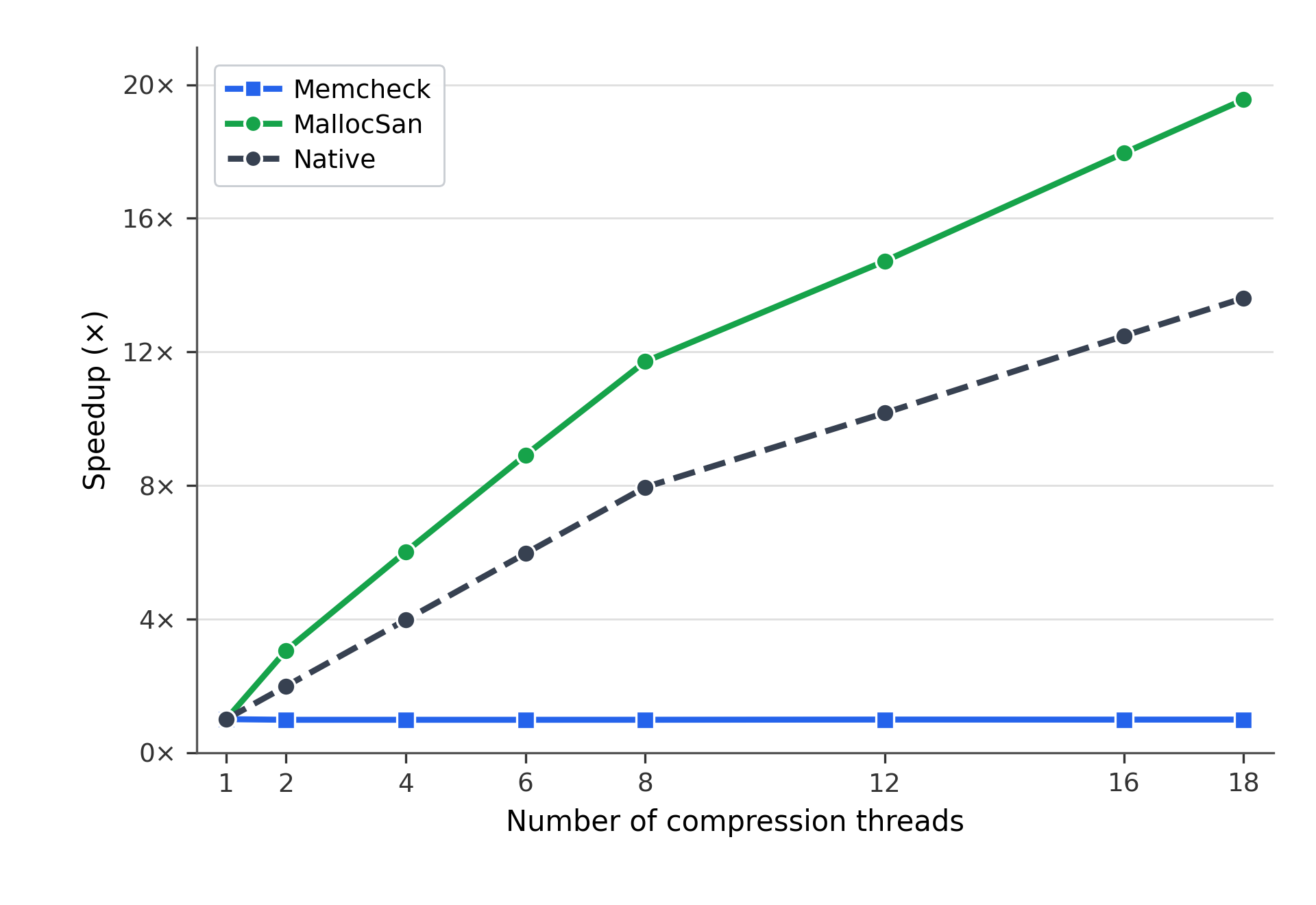}
    \par\smallskip\textbf{(d)} pigz: speedup
  \end{minipage}
  \caption{Multithreaded scaling of Native, MallocSan, and Memcheck on (a)--(b) the OpenMP-enabled SPEC CPU~2017 benchmark 644.nab\_s and (c)--(d) pigz. Thread counts refer to OpenMP application threads for nab and compression threads for pigz. Panels (a) and (c) show absolute execution time, where lower is better; panels (b) and (d) show speedup relative to each configuration's own one-thread execution, where higher is better.}
  \Description{Four line charts compare Native, MallocSan, and Memcheck across OpenMP thread counts for 644.nab_s and compression-thread counts for pigz. Native and MallocSan execution times decrease with additional threads while Memcheck remains nearly flat. Native and MallocSan have nearly identical speedups on nab. On pigz, MallocSan has a higher speedup relative to its own one-thread baseline but remains slower than Native in absolute time.}
  \label{fig:threads-scaling}
\end{figure*}

This normalized view compares detector overhead at each thread count but does not directly measure parallel speedup. Figures~\ref{fig:threads-scaling}(a) and~\ref{fig:threads-scaling}(c) therefore report mean absolute execution times, while Figures~\ref{fig:threads-scaling}(b) and~\ref{fig:threads-scaling}(d) report the speedup of each configuration relative to its own one-thread execution:

\[
S_c(T)=\frac{\overline{t}_c(1)}
{\overline{t}_c(T)},
\qquad
c \in \{\mathrm{Native},\mathrm{MallocSan},\mathrm{Memcheck}\},
\]

where $\overline{t}_c(T)$ is the mean execution time of configuration $c$ at
thread count $T$ for the workload under consideration. This metric measures
each configuration's improvement over its own one-thread baseline.
Consequently, overlapping speedup curves indicate comparable scaling rather
than equal absolute execution time; the absolute-time and normalized-overhead
panels show the detector cost.

\paragraph{OpenMP 644.nab\_s.}
MallocSan remains near Native throughout the sweep, with its normalized factor
rounding to $1.00\times$ at 18 threads. Memcheck begins at $17.6\times$ Native
with one thread and reaches $184.4\times$ with 18 threads. In absolute terms,
Native's mean execution time decreases from 6746.31~s to 652.70~s, while
MallocSan's decreases from 6850.87~s to 653.98~s. Their nearly overlapping
curves yield 18-thread speedups of $10.34\times$ and $10.48\times$,
respectively. Using $E_c(T)=S_c(T)/T$ to define parallel efficiency gives
$57.4\%$ for Native and $58.2\%$ for MallocSan. Memcheck's execution time
remains approximately constant, so its increasing normalized factor reflects
the scaling of the Native denominator rather than a comparable increase in
Memcheck's absolute cost.

\paragraph{Pigz compression.}
Native's mean execution time falls from 62.368~s with one compression thread
to 4.582~s with 18, a $13.61\times$ speedup. MallocSan falls from 107.166~s to
5.480~s, a higher one-thread-relative speedup of $19.56\times$. Its normalized
execution-time factor drops from $1.72\times$ at one compression thread to
$1.12\times$ at two, then increases gradually to $1.20\times$ at 18; across
the multithreaded settings, it stays within $1.12$--$1.20\times$ Native.
Thus, the higher relative speedup does not mean that MallocSan outperforms
Native in absolute time: at 18 compression threads it still takes about
$19.6\%$ longer. Memcheck's mean time changes only from 644.416~s to
652.500~s, yielding a $0.99\times$ speedup. Its normalized factor consequently
grows from $10.3\times$ to $142.4\times$ as Native becomes faster.

\paragraph{Patched-site intensity.}
The near overlap of Native and MallocSan in
Figures~\ref{fig:threads-scaling}(a)--(b) is consistent with 644.nab\_s's low
dynamic checking intensity relative to its long execution time. In the
one-thread profiles, all 271 NAB sites and 122 of the 123 pigz sites used the
same direct \texttt{JUMP} mechanism; the remaining pigz site used emulation
and accounted for only 7{,}630 hits. The profiles record 15.48 million total
site hits for NAB and 63.42 million for pigz; relative to their mean MallocSan
execution times of 6850.87~s and 107.166~s, respectively, these totals
correspond to approximately 2.26 thousand and 591.8 thousand hits per second,
a roughly $262\times$ higher dynamic rate for pigz. Despite their nearly
identical patch strategies, pigz therefore exercises MallocSan's recurring
checking path much more intensively, making its detector overhead visible while
the NAB curves nearly overlap.

MallocSan's higher speedup relative to its one-thread pigz baseline is a
separate effect, consistent with overlap between first-fault initialization
and useful compression work. As described in Section~\ref{sec:design}, when a
thread first faults at an uninitialized
instruction, it decodes the instruction, creates its shared metadata, and
requests patch installation. Threads that concurrently fault at the same
site wait for initialization to complete and then reuse the published
metadata and installed patch. Meanwhile, other compression workers can
execute already-patched sites or initialize distinct sites concurrently.
Patch installation itself remains serialized through the dedicated patch
worker. Increasing the compression-thread count can therefore distribute
one-time discovery and metadata preparation across workers and overlap
analysis and patching with useful execution. This provides a mechanism for
reducing the relative impact of initialization, although the aggregate
timings do not isolate its contribution from other scaling effects. Because
MallocSan's one-thread baseline includes a larger relative detector cost,
its $19.56\times$ speedup can exceed the configured count of 18 compression
threads without implying superlinear scaling of compression work itself.

\paragraph{Answer to RQ4.}
MallocSan preserves substantial parallel scaling on both the OpenMP
benchmark and the general-purpose compression application. On 644.nab\_s,
its 18-thread speedup ($10.48\times$) closely matches Native's
($10.34\times$), with a normalized execution-time factor near $1.0\times$.
On pigz, it achieves a $19.56\times$ speedup relative to its own one-thread
baseline, versus Native's $13.61\times$, while retaining a $1.20\times$
execution-time factor at 18 compression threads. Memcheck shows essentially
no parallel speedup on either workload. These results cover the two evaluated
workloads, inputs, and thread ranges on this machine; they do not establish
the same behavior for all multithreaded applications or hardware topologies.

\section{General Discussion}\label{sec:discussion}

This section interprets the evaluation results and delineates the conditions under which MallocSan's guarantees apply.

\subsection{Evaluation Findings and Design Tradeoffs}\label{sec:discussion-findings}

Within the evaluated scope, the accuracy results support the correctness of MallocSan's OID-based interception and bounds-validation mechanisms across the selected control- and data-flow variants. They do not establish complete memory-safety coverage: the excluded stack and intra-object violations illustrate the boundary of allocation-granularity heap protection, while the synthetic nature of Juliet limits generalization to more complex defects.

The comparison with Memcheck highlights the tradeoff at the center of MallocSan's deployment model and explains why the tools' costs diverge. Memcheck translates executed code and propagates shadow state to perform broad addressability checking across the process, regardless of whether an access involves an object of current interest. MallocSan instead concentrates its checking cost on instructions that dereference protected heap pointers. When a site supports direct \texttt{JUMP}-based redirection, recurring checks remain in userspace, while post-handler elimination and deferral reduce additional handler work and control transfers. This difference in scope explains the execution-time divergence across workloads. Because the experiment protects every eligible heap allocation, it represents MallocSan's widest OID-tagging policy; a narrower policy could reduce overhead by protecting only allocations relevant to the analysis.

The behavior of 605.mcf\_s and 619.lbm\_s exposes distinct unfavorable cases for this model. MallocSan's deferral analysis conservatively stops at nearly all control-flow boundaries, which fragments deferral windows, restores protected pointers relatively early, and requires subsequent accesses to re-enter the checking path. In 605.mcf\_s, dense control flow repeatedly shortens these windows. In 619.lbm\_s, a streaming lattice-Boltzmann kernel executes a dense set of memory accesses at high dynamic frequency in hot loops, so a loop boundary or another eligibility condition that limits deferral across the kernel leaves even a small set of protected sites costly. In either workload, a hot site that requires the \texttt{TRAP} fallback adds signal delivery, handler dispatch, and context recovery on every invocation. The results are therefore consistent with an interaction between limited deferral and frequently executed fault-driven sites rather than with a uniform per-access cost across workloads.

The profile-guided study shows that replacing eligible hot sites with direct \texttt{JUMP}-based redirection can recover some of this fallback cost. The residual overhead after the transformation exposes the limits of this optimization. Sites without a safe rewrite window remain on the \texttt{TRAP} path, while rewritten sites still perform OID stripping, bounds validation, and tag restoration. The attainable benefit consequently depends on the dynamic coverage of eligible sites and the residual cost of checking along the rewritten path.

The memory pattern similarly follows from the tools' memory-management designs. Memcheck maintains shadow metadata, allocation redzones, and runtime translation structures. MallocSan's persistent state includes a bounded object table, an instruction table that grows with the number of distinct protected sites, and fixed-capacity trampoline and OLX pools. The relative effect of these largely additive costs depends on workload scale: the runtime state is prominent relative to 648.exchange2\_s's exceptionally small Native footprint but contributes little in relative terms on workloads with large Native footprints.

The multithreaded results on 644.nab\_s and pigz reflect the tools' execution
models. Memcheck serializes client-thread execution, preventing the
instrumented application from exploiting multiple
cores~\cite{nethercote2007valgrind}; its nearly constant execution time on both
workloads contrasts with Native's decreasing time, causing the normalized
execution-time ratio to widen. MallocSan does not serialize ordinary
application execution: application threads run concurrently, faults are
handled on the faulting thread's alternate signal stack, and steady-state
checking uses thread-local execution state. Because \texttt{libpatch} is not
thread-safe, patch installation is routed through a dedicated worker
(Section~\ref{sec:design}), but this serialization occurs once per newly
encountered instruction site rather than on every protected access.

The two workloads expose different aspects of this design. On 644.nab\_s,
MallocSan's execution-time and speedup curves closely match Native's. Pigz
extends the evidence to a general-purpose compression application, but also
illustrates why relative speedup must be distinguished from absolute cost:
at 18 compression threads, MallocSan achieves a $19.56\times$ speedup against
its own one-thread baseline, versus Native's $13.61\times$, while still
taking $1.20\times$ as long. As discussed in Section~\ref{sec:eval-scaling},
this higher relative speedup is consistent with overlap between first-fault
initialization and useful execution. Workers can analyze distinct newly
faulting sites concurrently, while workers reaching the same site wait for
initialization and then reuse its metadata and patch. Although installation
remains serialized, analysis and patching can overlap with other workers'
compression work, reducing the relative impact of initialization.

Scaling must also be interpreted against the processor's heterogeneous
topology: its performance cores support simultaneous multithreading, while
its efficiency cores offer different throughput, so additional workers do
not necessarily contribute the capacity of independent performance cores.
Native execution on the same hardware and input therefore provides the
relevant empirical baseline. Equal
configured thread counts do not imply identical total software-thread counts:
for pigz they count compression workers, excluding I/O and detector helper
threads. Moreover, speedup relative to an instrumented one-thread baseline
includes changes in detector overhead as well as application parallelism;
MallocSan's speedup exceeding the configured pigz compression-thread count
does not imply superlinear scaling of compression itself. The absolute-time
and speedup results
support preserved parallelism on both workloads, with workload-dependent
detector costs.

Taken together, these results place MallocSan and Memcheck at different operating points rather than making them direct substitutes. Memcheck is better suited to analyses that require broad whole-process memory checking and can tolerate substantial overhead. MallocSan is designed for unmodified binaries in settings where deterministic per-object validation of selected heap accesses must coexist with lower overhead and concurrent application execution. Its protection policy makes this tradeoff explicit: narrowing the policy reduces both checking cost and coverage, whereas broadening it extends heap coverage at a correspondingly higher cost.

Within this niche, MallocSan is most applicable to dynamically linked x86-64
Linux applications whose protected allocations pass through supported
allocation interfaces, whose pointers do not repurpose MallocSan's tag bits,
and whose protection requirements align with allocation-granularity heap
checking. We do not estimate how common this application class is: the
evaluated workloads were selected for relevance and compatibility rather than
sampled from a representative application population. Quantifying its
prevalence would require a corpus study measuring execution compatibility,
eligible-allocation coverage, and protected-access coverage across a defined
set of real applications.

\subsection{Limitations and Threats to Validity}\label{sec:limitations}
MallocSan relies on assumptions about the virtual-address layout, allocation interface, and application use of pointer bits. These assumptions hold for many dynamically linked x86-64 Linux applications, but not universally. When they do not hold, MallocSan may protect fewer objects, omit allocations from coverage, or be incompatible with the application.

\paragraph{Reduced tag space under five-level paging.}
Under conventional four-level paging (LA48), lower-half user virtual addresses use bits~0--46, while bits~47--63 are zero in canonical addresses. MallocSan can therefore store a 15-bit OID in bits~48--62 while keeping bit~63 clear; any nonzero OID makes the resulting pointer noncanonical. On systems actively using five-level paging (LA57), lower-half user addresses may use bits~0--55. To preserve the noncanonical-address trap, the tag must be restricted to bits~56--62, reducing the number of simultaneously available nonzero OIDs from 32{,}767 to 127.

\paragraph{Non-standard allocators.}
MallocSan discovers heap objects by interposing standard allocation functions through \texttt{LD\_PRELOAD}. Allocations that are not returned through these interposed entry points receive neither an OID nor an object-table entry. This includes objects managed by custom arenas, slab allocators, or memory pools that obtain or subdivide storage without using the intercepted interface, as well as allocations obtained directly through \texttt{mmap} or \texttt{brk}. Errors within such objects may therefore go undetected. Stack-allocated buffers and global variables are likewise outside the scope of MallocSan's heap-oriented protection model.

\paragraph{Applications that repurpose high pointer bits.}
MallocSan assumes that the application does not use the selected tag bits for its own metadata. This assumption fails for programs or runtimes that interpret the high bits of pointers. One example is GCC's garbage-collected heap (\texttt{ggc\_alloc}/\texttt{ggc\_free}), whose runtime uses the raw pointer value, including its high bits, to index an internal page table. A MallocSan tag can therefore cause the pointer to be interpreted as referring to a different page-table entry, resulting in an invalid lookup or a crash. Applications with similar pointer-encoding conventions require allocator-specific integration or an alternative tag representation and are not supported by the current implementation.

\paragraph{Scope of scaling evidence.}
The scaling experiments cover two workloads on one heterogeneous processor,
with the inputs and settings reported in Section~\ref{sec:eval-scaling}.
They do not establish the same behavior for other workload classes,
compression inputs or levels, or hardware topologies. Aggregate timings also
do not isolate the contribution of concurrent first-fault analysis and its
overlap with useful work; quantifying that contribution would require
targeted profiling or an ablation study.

\section{Related Work} \label{sec:relwork}
This section surveys userspace memory-safety techniques for C and C++ applications, focusing on the design dimensions most relevant to MallocSan. Kernel-level mechanisms, managed-runtime defenses, and memory-safe language migration strategies fall outside our scope.

% ---------------------------------------------------------------
\subsection{Static Detection of Memory Errors} \label{sec:static}

Static analysis imposes no runtime overhead on the analyzed program, but practical analyzers must balance soundness, precision, and scalability~\cite{shahriar2010classification}. Lightweight tools such as Flawfinder and Cppcheck scale to large codebases but may miss defects requiring deeper semantic or interprocedural reasoning~\cite{wheeler2021flawfinder,daniel2026cppcheck}. Path-sensitive tools such as ARCHER and Marple reason more precisely about feasible executions but must contend with path explosion~\cite{xie2003archer,le2008marple}. Empirical studies nevertheless reveal substantial limitations. On the Toyota ITC benchmark, the best-performing evaluated configuration achieved a robust detection rate of only 25.67\%~\cite{arusoaie2017comparison}. Similarly, among 159 real-world buffer-overflow vulnerabilities, only 22 fixes (13.84\%) changed the alerts produced by Cppcheck or Flawfinder~\cite{pereira2021characterizing}. Symbolic execution, exemplified by KLEE, can uncover deeper errors through path-constraint solving, but exponential state growth and symbolic-memory challenges limit its scalability; compositional summaries mitigate rather than eliminate these problems~\cite{cadar2008klee,godefroid2007compositional}.

Learning-based detectors seek to infer implicit vulnerability signatures. Token-based approaches include a neural memory network that predicts buffer overruns directly from raw C/C++ source and a CNN-BLSTM model augmented with static taint analysis for CWE-119 and CWE-399~\cite{choi2017EndtoendPrediction,niu2020deep}. Graph-based approaches encode abstract syntax, control flow, and data flow. One graph-and-transformer model reportedly outperformed Cppcheck and Flawfinder on a dataset of more than 12,000 code fragments, while related work developed graph-based data-preparation pipelines for CI/CD environments~\cite{savenko2025detection,savenko2025graphbased}. However, source-level models require parseable code, while learning-based detectors depend on accurately labeled data and may generalize poorly to projects that differ from their training distributions. Assembly-level representations reduce dependence on source availability, but their generalization to real-world binaries remains insufficiently established~\cite{dahl2020stackbased}. More generally, false positives increase analyst workload and impede operational adoption~\cite{johnson2013why}. Like conventional static analyzers, these methods detect rather than prevent vulnerabilities and therefore provide no runtime protection when defects are missed.

% ---------------------------------------------------------------
\subsection{Compiler-Based Dynamic Enforcement} \label{sec:dynamic}

Compiler-based sanitizers are widely used for dynamic memory-error testing. ASan maps every eight bytes of application memory to one shadow byte, places redzones around objects, and quarantines freed heap memory to detect spatial errors and heap use-after-free, with a reported average slowdown of 73\% and a memory overhead of approximately $3.4\times$~\cite{serebryany2012addresssanitizer}. MemorySanitizer similarly maintains bit-precise shadow state to detect uses of uninitialized values~\cite{stepanov2015memorysanitizer,vanoorschot2023memory}. CMASan extends ASan to objects managed by custom allocators by identifying allocator functions and instrumenting their APIs, although categorizing allocator API families still requires user input~\cite{hong2025cmasan}. By contrast, StackGuard is a runtime mitigation that uses compiler-inserted canaries to protect return addresses rather than general object bounds~\cite{cowan1998stackguard}.

The Memory Sanitizer Evaluation Tool (MSET) distinguishes theoretical guarantees from realized detection coverage: location-based approaches using redzones or guard pages can miss spatial errors, whereas identity-based per-pointer tracking can support deterministic object-bounds checking~\cite{vintila2025evaluating}. SoftBound associates base-and-bound metadata with each pointer, while CETS adds key-and-lock metadata and changes the value stored at the lock upon deallocation, causing checks on stale pointers to fail~\cite{nagarakatte2009softbound,nagarakatte2010cets}. The latest software-only SoftBoundCETS prototype, based on LLVM~17.6, averages 72\% overhead across a collection of SPEC benchmarks~\cite{nagarakatte2024full}. EffectiveSan uses low-fat allocation and dynamic type metadata to check object and sub-object bounds and detect type confusion, although its temporal-error coverage is partial~\cite{duck2018effectivesan}. These approaches rely on compiler instrumentation and therefore cannot fully protect prebuilt binaries, even when interoperability with uninstrumented code is supported.

% ---------------------------------------------------------------
\subsection{Hardware-Assisted Detection} \label{sec:hardware}

Hardware assistance reduces enforcement cost but introduces platform dependencies. CHERI extends the ISA with 128-bit capability registers encoding base, length, permissions, and a hardware-protected validity tag, enabling checks at single-digit percentage overhead~\cite{watson2015cheri,woodruff2019cheri}. However, it requires CHERI processors, a CHERI-aware OS such as CheriBSD, and recompilation of the software stack for a capability ABI~\cite{sri2026cheribsd}. No-FAT derives bounds from a binning allocator's fixed-size regions using a hardware table and three new instructions, adding 8\% overhead on SPEC CPU~2017~\cite{ibnziad2021nofat}; PACMem embeds object metadata in ARM pointer signatures, providing spatial and temporal coverage at 69\% overhead~\cite{li2022pacmem}.

On AArch64, compiler-based HWASan assigns 8-bit tags to 16-byte granules and embeds matching tags in pointers' top bytes, incurring about $2\times$ overhead with a 0.39\% collision-based false-negative rate~\cite{serebryany2018memory}. MTSan instead uses ARM Memory Tagging Extension to instrument COTS binaries and probabilistically infer heap, stack, and global boundaries without source code. Its 4-bit tags reduce overhead to about $1.82\times$ but increase the irreducible false-negative rate to 6.25\%~\cite{chen2023mtsan}. These approaches therefore depend on a capability architecture, custom ISA extensions, or ARM-specific pointer and tagging features. None readily protects commodity x86-64 binaries without recompilation or execution-platform changes.

% ---------------------------------------------------------------
\subsection{Binary-Level Instrumentation} \label{sec:binary}

Without source code, tools rely on dynamic binary translation (DBT) or static rewriting. Valgrind's Memcheck instruments every load and store during translation while maintaining shadow state. Because stripped binaries lack precise stack and global boundaries, its object-level checking focuses on the heap; its translation pipeline also imposes a $17$--$40\times$ slowdown, limiting continuous monitoring and fuzzing~\cite{nethercote2007valgrind}. RetroWrite offers a lower-cost alternative by reconstructing assembly from position-independent ELF binaries and injecting ASan-compatible checks at ASan-like overhead. However, it requires sufficient relocation information and well-structured compiler output, and provides limited stack and global coverage~\cite{dinesh2020retrowrite}.

MTSan combines static rewriting with MTE to cover heap, stack, and global objects in COTS AArch64 binaries. It detected 18 of 27 real vulnerabilities, compared with 12 for ASan-RetroWrite and 13 for Memcheck, largely because of its stack and global coverage~\cite{chen2023mtsan}. However, it cannot automatically instrument custom allocators that bypass \texttt{malloc}/\texttt{free}, while its 16-byte granularity misses sub-object overflows. Together, these systems expose the binary-level semantic gap: recovering object boundaries, fitting metadata into fixed layouts, and covering non-standard allocation paths remain difficult. Existing designs bridge this gap only partially through whole-program translation, relocation-driven reconstruction, probabilistic inference, or platform-specific hardware.

% ---------------------------------------------------------------
\subsection{Runtime Interposition and Allocator-Based Defenses} \label{sec:interposition}

Runtime interposition replaces allocator functions through the dynamic linker, enabling protection of prebuilt binaries without source access, binary rewriting, or specialized hardware. DieHard approximates an infinite heap by randomizing object placement and delaying reuse, making adjacent overflows and use-after-free less likely to affect live objects. It averages 8--12\% overhead on standard benchmarks, but nonlinear overflows into live objects and unlucky reuse remain undetected~\cite{berger2006diehard}. FreeGuard combines randomized chunk placement, canaries, and guard pages at below 2\% average overhead. Although its bounds protection remains probabilistic, it deterministically detects double and invalid frees and reports precise call stacks~\cite{silvestro2017freeguard}.

MarkUs provides deterministic temporal safety by delaying reuse until a conservative background scan finds no live pointers to the freed object, but performs no bounds checking~\cite{ainsworth2020markus}. CRCount maintains reference counts and nullifies pointers to freed objects, but requires compiler instrumentation to propagate pointer tracking~\cite{shin2019crcount}. These systems demonstrate that interposition is practical for heap protection in existing binaries, yet allocator-level visibility cannot validate individual accesses. They provide probabilistic spatial protection (DieHard and FreeGuard), temporal-only protection (MarkUs), or compiler-dependent pointer tracking (CRCount); none offers identity-based deterministic spatial detection for heap objects in unmodified binaries.

MallocSan fills this gap. Like DieHard and FreeGuard, it uses \texttt{LD\_PRELOAD} on commodity x86-64 systems without source access, recompilation, or specialized hardware. Unlike them, it validates every protected access against per-object bounds, providing deterministic, identity-based spatial checking previously limited to source-level or hardware-assisted approaches, plus policy-driven use-after-free reporting (Section~\ref{sec:design}). This comparison concerns the guarantee for objects selected by MallocSan's protection policy; allocations that bypass the interposed interface remain outside its scope. Runtime binary patching and optional profile-guided rewriting reduce the cost at hot instruction sites.

\section{Conclusion}\label{sec:conclusion}
This paper presented MallocSan, a heap sanitizer for deterministic, per-object spatial checking of unmodified x86-64 Linux binaries. MallocSan interposes allocation routines, assigns each protected allocation an object identifier (OID), and encodes it in otherwise unused pointer bits. The first execution of an unpatched memory-access instruction through a tagged pointer triggers a noncanonical-address fault, allowing the runtime to validate the access and, when possible, install a direct \texttt{JUMP}-based redirection for later executions. \texttt{TRAP}-based patching, in-handler emulation, and single-stepping preserve checking at sites that cannot use direct redirection. Together, these mechanisms provide policy-scoped heap protection and optional use-after-free reporting without source access, recompilation, or specialized hardware.

% PIGZ UPDATE: The conclusion now summarizes both RQ4 workloads.
The evaluation supports the model's correctness within its stated scope while showing workload-dependent performance. Across the seven selected SPEC CPU~2017 benchmarks, MallocSan's execution-time geometric-mean factor was $4.82\times$, compared with $18.55\times$ for Memcheck. On the in-scope Juliet tests, MallocSan detected every manifested violation and issued no reports on the good executions. MallocSan preserved substantial parallel scaling on both evaluated multithreaded workloads. At 18 threads on 644.nab\_s, its $10.48\times$ speedup closely matched Native's $10.34\times$ speedup, with a normalized execution-time factor near $1.0\times$. At 18 compression threads on pigz, MallocSan achieved a $19.56\times$ speedup relative to its own one-thread baseline, compared with Native's $13.61\times$, while taking $1.20\times$ as long. Memcheck showed essentially no parallel speedup on either workload.

Taken together, the results show that runtime interposition and identity-based spatial checking can coexist in a practical sanitizer for native binaries, subject to a defined scope. MallocSan protects heap allocations obtained through interposed interfaces on x86-64 Linux. Stack and global objects, allocations managed through unsupported custom interfaces, and applications that repurpose the selected pointer bits fall outside this scope. Future work should extend direct redirection to sites that currently remain unpatchable, broaden instruction coverage in in-handler emulation, and reduce the recurring cost of the patched checking path. Support is also needed for platforms where features such as ARM Top-Byte Ignore and Intel Linear Address Masking prevent tagged pointers from producing MallocSan's address faults. Such platforms would require explicit checks or another interception mechanism, potentially through binary rewriting analogous in function to HWASan's compiler-inserted checks~\cite{serebryany2018memory}. These extensions would broaden MallocSan's applicability while preserving its central objective: deterministic validation of protected heap accesses in binaries that cannot be rebuilt.

%-----------------------

\section*{Data and Artifact Availability}
To facilitate reproducibility, the source code of MallocSan and of our extended version of \texttt{libpatch} is publicly available at \url{https://github.com/adel-belkhiri/MallocSan} and \url{https://github.com/adel-belkhiri/libpatch}, respectively.

\section*{Acknowledgments}
We thank Ciena, Ericsson, EfficiOS, and the Natural Sciences and Engineering Research Council of Canada for supporting this work.

\bibliographystyle{IEEEtran}
%\balance
\bibliography{references}

@misc{sri2026cheribsd,
  author = {{SRI International} and {University of Cambridge}},
  title  = {{CheriBSD}},
  howpublished = {\url{https://www.cheribsd.org/}},
  year   = {2026}
}

@misc{daniel2026cppcheck,
  author = {Marjamäki, Daniel and Team, Cppcheck},
  title  = {Cppcheck 2.20},
  howpublished = {\url{https://cppcheck.sourceforge.io/}},
  year   = {2026}
}

@misc{wheeler2021flawfinder,
  author = {Wheeler, David A.},
  title  = {Flawfinder 2.0.19},
  howpublished = {\url{https://dwheeler.com/flawfinder/}},
  month  = aug,
  year   = {2021}
}

@article{niu2020deep,
  title    = {A deep learning based static taint analysis approach for {IoT} software vulnerability location},
  volume   = {152},
  issn     = {0263-2241},
  url      = {https://www.sciencedirect.com/science/article/pii/S026322411931005X},
  doi      = {10.1016/j.measurement.2019.107139},
  urldate  = {2026-04-02},
  journal  = {Measurement},
  author   = {Niu, Weina and Zhang, Xiaosong and Du, Xiaojiang and Zhao, Lingyuan and Cao, Rong and Guizani, Mohsen},
  month    = feb,
  year     = {2020},
  pages    = {107139}
}

@inproceedings{le2008marple,
  address    = {Atlanta Georgia},
  title      = {Marple: a demand-driven path-sensitive buffer overflow detector},
  isbn       = {978-1-59593-995-1},
  shorttitle = {Marple},
  url        = {https://dl.acm.org/doi/10.1145/1453101.1453137},
  doi        = {10.1145/1453101.1453137},
  language   = {en},
  urldate    = {2026-04-02},
  booktitle  = {Proceedings of the 16th {ACM} {SIGSOFT} {International} {Symposium} on {Foundations} of software engineering},
  publisher  = {ACM},
  author     = {Le, Wei and Soffa, Mary Lou},
  month      = nov,
  year       = {2008},
  pages      = {272--282}
}

@inproceedings{xie2003archer,
  address    = {Helsinki Finland},
  title      = {{ARCHER}: using symbolic, path-sensitive analysis to detect memory access errors},
  isbn       = {978-1-58113-743-9},
  shorttitle = {{ARCHER}},
  url        = {https://dl.acm.org/doi/10.1145/940071.940115},
  doi        = {10.1145/940071.940115},
  language   = {en},
  urldate    = {2026-04-02},
  booktitle  = {Proceedings of the 9th {European} software engineering conference held jointly with 11th {ACM} {SIGSOFT} international symposium on {Foundations} of software engineering},
  publisher  = {ACM},
  author     = {Xie, Yichen and Chou, Andy and Engler, Dawson},
  month      = sep,
  year       = {2003},
  pages      = {327--336}
}

@inproceedings{savenko2025graphbased,
  title     = {Graph-based data preparation for detecting buffer overflow vulnerabilities in code within {CI}/{CD} pipelines},
  booktitle = {{AdvAIT}-2025: 2nd {International} {Workshop} on {Advanced} {Applied} {Information} {Technologies}},
  author    = {Savenko, Oleg and Lips, Silvia and Gaj, Piotr and Sierhieiev, Yevhenii},
  year      = {2025}
}

@inproceedings{savenko2025detection,
  title     = {Detection of buffer overflow vulnerabilities in system software based on a graph and transformer model},
  booktitle = {The {International} {Workshop} on {Applied} {Intelligent} {Security} {Systems} in {Law} {Enforcement}},
  author    = {Savenko, Oleg and Gaj, Piotr and Sierhieiev, Yevhenii},
  year      = {2025}
}

@inproceedings{chen2023mtsan,
  title      = {{MTSan}: {A} {Feasible} and {Practical} {Memory} {Sanitizer} for {Fuzzing} {COTS} {Binaries}},
  isbn       = {978-1-939133-37-3},
  shorttitle = {{MTSan}},
  url        = {https://www.usenix.org/conference/usenixsecurity23/presentation/chen-xingman},
  language   = {en},
  urldate    = {2026-03-17},
  booktitle  = {32nd {USENIX} {Security} {Symposium} ({USENIX} {Security} 23)},
  author     = {Chen, Xingman and Shi, Yinghao and Jiang, Zheyu and Li, Yuan and Wang, Ruoyu and Duan, Haixin and Wang, Haoyu and Zhang, Chao},
  year       = {2023},
  pages      = {841--858}
}

@inproceedings{shin2019crcount,
  address    = {San Diego, CA},
  title      = {{CRCount}: {Pointer} {Invalidation} with {Reference} {Counting} to {Mitigate} {Use}-after-free in {Legacy} {C}/{C}++},
  isbn       = {978-1-891562-55-6},
  shorttitle = {{CRCount}},
  url        = {https://www.ndss-symposium.org/wp-content/uploads/2019/02/ndss2019_05A-4_Shin_paper.pdf},
  doi        = {10.14722/ndss.2019.23541},
  language   = {en},
  urldate    = {2026-03-27},
  booktitle  = {Proceedings 2019 {Network} and {Distributed} {System} {Security} {Symposium}},
  publisher  = {Internet Society},
  author     = {Shin, Jangseop and Kwon, Donghyun and Seo, Jiwon and Cho, Yeongpil and Paek, Yunheung},
  year       = {2019}
}

@inproceedings{dinesh2020retrowrite,
  address    = {San Francisco, CA, USA},
  title      = {{RetroWrite}: {Statically} {Instrumenting} {COTS} {Binaries} for {Fuzzing} and {Sanitization}},
  copyright  = {https://ieeexplore.ieee.org/Xplorehelp/downloads/license-information/IEEE.html},
  isbn       = {978-1-7281-3497-0},
  shorttitle = {{RetroWrite}},
  url        = {https://ieeexplore.ieee.org/document/9152762/},
  doi        = {10.1109/SP40000.2020.00009},
  urldate    = {2026-03-31},
  booktitle  = {2020 {IEEE} {Symposium} on {Security} and {Privacy} ({SP})},
  publisher  = {IEEE},
  author     = {Dinesh, Sushant and Burow, Nathan and Xu, Dongyan and Payer, Mathias},
  month      = may,
  year       = {2020},
  pages      = {1497--1511}
}

@inproceedings{arusoaie2017comparison,
  title     = {A comparison of open-source static analysis tools for vulnerability detection in {C/C++} code},
  booktitle = {2017 19th {International} {Symposium} on {Symbolic} and {Numeric} {Algorithms} for {Scientific} {Computing} ({SYNASC})},
  publisher = {IEEE},
  author    = {Arusoaie, Andrei and Ciobâca, Stefan and Craciun, Vlad and Gavrilut, Dragos and Lucanu, Dorel},
  year      = {2017},
  pages     = {161--168}
}

@inproceedings{berger2006diehard,
  address    = {Ottawa Ontario Canada},
  title      = {{DieHard}: probabilistic memory safety for unsafe languages},
  isbn       = {978-1-59593-320-1},
  shorttitle = {{DieHard}},
  url        = {https://dl.acm.org/doi/10.1145/1133981.1134000},
  doi        = {10.1145/1133981.1134000},
  language   = {en},
  urldate    = {2026-03-27},
  booktitle  = {Proceedings of the 27th {ACM} {SIGPLAN} {Conference} on {Programming} {Language} {Design} and {Implementation}},
  publisher  = {ACM},
  author     = {Berger, Emery D. and Zorn, Benjamin G.},
  month      = jun,
  year       = {2006},
  pages      = {158--168}
}

@inproceedings{cadar2008klee,
  address   = {San Diego, California},
  series    = {{OSDI}'08},
  title     = {{KLEE}: unassisted and automatic generation of high-coverage tests for complex systems programs},
  booktitle = {Proceedings of the 8th {USENIX} conference on operating systems design and implementation},
  publisher = {USENIX Association},
  author    = {Cadar, Cristian and Dunbar, Daniel and Engler, Dawson},
  year      = {2008},
  pages     = {209--224}
}

@misc{choi2017EndtoendPrediction,
  title   = {End-to-end prediction of buffer overruns from raw source code via neural memory networks},
  howpublished = {arXiv:1703.02458},
  author  = {Choi, Min-je and Jeong, Sehun and Oh, Hakjoo and Choo, Jaegul},
  year    = {2017}
}

@inproceedings{cowan1998stackguard,
  title     = {Stackguard: {Automatic} adaptive detection and prevention of buffer-overflow attacks},
  volume    = {98},
  booktitle = {{USENIX} security symposium},
  publisher = {San Antonio, TX},
  author    = {Cowan, Crispan and Pu, Calton and Maier, Dave and Walpole, Jonathan and Bakke, Peat and Beattie, Steve and Grier, Aaron and Wagle, Perry and Zhang, Qian and Hinton, Heather},
  year      = {1998},
  pages     = {63--78}
}

@misc{dahl2020stackbased,
  title   = {Stack-based buffer overflow detection using recurrent neural networks},
  howpublished = {arXiv:2012.15116},
  author  = {Dahl, William Arild and Erdodi, Laszlo and Zennaro, Fabio Massimo},
  year    = {2020}
}

@phdthesis{dorostkar2026addressmonitor,
  title        = {{D\'etection \`a faible surco\^ut des conditions de course et des violations de la s\^uret\'e de la m\'emoire de tas}},
  author       = {Dorostkar, Farzam},
  year         = 2025,
  journal      = {ProQuest Dissertations and Theses},
  address      = {Canada -- Quebec, CA},
  collaborator = {Dagenais, Michel and Li, Heng and Boyer, Fran{\c c}ois-Raymond},
  isbn         = {9798277489109},
  langid       = {french},
  school       = {Ecole Polytechnique, Montreal (Canada)}
}

@misc{libtiff_issue435,
  author       = {{LibTIFF Project}},
  title        = {{tiffcrop}: heap-buffer-overflow in
                  {extractContigSamplesShifted24bits}, {tiffcrop.c:3604}},
  year         = {2022},
  month        = jun,
  howpublished = {GitLab Issue \#435,
                  \url{https://gitlab.com/libtiff/libtiff/-/issues/435}},
  note         = {Reported June 15, 2022}
}

@misc{nvd_cve20223598,
  author       = {{National Institute of Standards and Technology}},
  title        = {{CVE-2022-3598}},
  year         = {2022},
  month        = oct,
  howpublished = {National Vulnerability Database,
                  \url{https://nvd.nist.gov/vuln/detail/CVE-2022-3598}}
}

@inproceedings{duck2020binary,
  address   = {London UK},
  title     = {Binary rewriting without control flow recovery},
  isbn      = {978-1-4503-7613-6},
  url       = {https://dl.acm.org/doi/10.1145/3385412.3385972},
  doi       = {10.1145/3385412.3385972},
  language  = {en},
  urldate   = {2026-03-11},
  booktitle = {Proceedings of the 41st {ACM} {SIGPLAN} {Conference} on {Programming} {Language} {Design} and {Implementation}},
  publisher = {ACM},
  author    = {Duck, Gregory J. and Gao, Xiang and Roychoudhury, Abhik},
  month     = jun,
  year      = {2020},
  pages     = {151--163}
}

@inproceedings{godefroid2007compositional,
  address   = {Nice France},
  title     = {Compositional dynamic test generation},
  isbn      = {978-1-59593-575-5},
  url       = {https://dl.acm.org/doi/10.1145/1190216.1190226},
  doi       = {10.1145/1190216.1190226},
  language  = {en},
  urldate   = {2026-03-27},
  booktitle = {Proceedings of the 34th annual {ACM} {SIGPLAN}-{SIGACT} symposium on {Principles} of programming languages},
  publisher = {ACM},
  author    = {Godefroid, Patrice},
  month     = jan,
  year      = {2007},
  pages     = {47--54}
}

@inproceedings{hong2025cmasan,
  title     = {{CMASan}: {Custom} memory allocator-aware address sanitizer},
  booktitle = {2025 {IEEE} {Symposium} on {Security} and {Privacy} ({SP})},
  publisher = {IEEE},
  author    = {Hong, Junwha and Jang, Wonil and Kim, Mijung and Yu, Lei and Kwon, Yonghwi and Jeon, Yuseok},
  year      = {2025},
  pages     = {740--757}
}

@inproceedings{ibnziad2021nofat,
  title      = {No-{FAT}: {Architectural} {Support} for {Low} {Overhead} {Memory} {Safety} {Checks}},
  issn       = {2575-713X},
  shorttitle = {No-{FAT}},
  url        = {https://ieeexplore.ieee.org/abstract/document/9499774},
  doi        = {10.1109/ISCA52012.2021.00076},
  urldate    = {2026-03-17},
  booktitle  = {2021 {ACM}/{IEEE} 48th {Annual} {International} {Symposium} on {Computer} {Architecture} ({ISCA})},
  author     = {Ibn Ziad, Mohamed Tarek and Arroyo, Miguel A. and Manzhosov, Evgeny and Piersma, Ryan and Sethumadhavan, Simha},
  month      = jun,
  year       = {2021},
  pages      = {916--929}
}

@inproceedings{johnson2013why,
  address   = {San Francisco, CA, USA},
  title     = {Why don't software developers use static analysis tools to find bugs?},
  isbn      = {978-1-4673-3076-3 978-1-4673-3073-2},
  url       = {http://ieeexplore.ieee.org/document/6606613/},
  doi       = {10.1109/ICSE.2013.6606613},
  urldate   = {2026-03-27},
  booktitle = {2013 35th {International} {Conference} on {Software} {Engineering} ({ICSE})},
  publisher = {IEEE},
  author    = {Johnson, Brittany and Song, Yoonki and Murphy-Hill, Emerson and Bowdidge, Robert},
  month     = may,
  year      = {2013},
  pages     = {672--681}
}

@inproceedings{lattner2004llvm,
  address    = {San Jose, CA, USA},
  title      = {{LLVM}: {A} compilation framework for lifelong program analysis \& transformation},
  isbn       = {978-0-7695-2102-2},
  shorttitle = {{LLVM}},
  url        = {http://ieeexplore.ieee.org/document/1281665/},
  doi        = {10.1109/CGO.2004.1281665},
  urldate    = {2026-03-27},
  booktitle  = {International {Symposium} on {Code} {Generation} and {Optimization}, 2004. {CGO} 2004.},
  publisher  = {IEEE},
  author     = {Lattner, C. and Adve, V.},
  year       = {2004},
  pages      = {75--86}
}

@inproceedings{li2022pacmem,
  address    = {New York, NY, USA},
  series     = {{CCS} '22},
  title      = {{PACMem}: {Enforcing} {Spatial} and {Temporal} {Memory} {Safety} via {ARM} {Pointer} {Authentication}},
  isbn       = {978-1-4503-9450-5},
  shorttitle = {{PACMem}},
  url        = {https://dl.acm.org/doi/10.1145/3548606.3560598},
  doi        = {10.1145/3548606.3560598},
  urldate    = {2026-03-17},
  booktitle  = {Proceedings of the 2022 {ACM} {SIGSAC} {Conference} on {Computer} and {Communications} {Security}},
  publisher  = {Association for Computing Machinery},
  author     = {Li, Yuan and Tan, Wende and Lv, Zhizheng and Yang, Songtao and Payer, Mathias and Liu, Ying and Zhang, Chao},
  month      = nov,
  year       = {2022},
  pages      = {1901--1915}
}

@article{nagarakatte2024full,
  title    = {Full {Spatial} and {Temporal} {Memory} {Safety} for {C}},
  volume   = {22},
  issn     = {1558-4046},
  url      = {https://ieeexplore.ieee.org/document/10439147/},
  doi      = {10.1109/MSEC.2024.3363142},
  number   = {4},
  urldate  = {2026-03-17},
  journal  = {IEEE Security \& Privacy},
  author   = {Nagarakatte, Santosh},
  month    = jul,
  year     = {2024},
  pages    = {30--39}
}

@inproceedings{nagarakatte2010cets,
  address    = {Toronto Ontario Canada},
  title      = {{CETS}: compiler enforced temporal safety for {C}},
  isbn       = {978-1-4503-0054-4},
  shorttitle = {{CETS}},
  url        = {https://dl.acm.org/doi/10.1145/1806651.1806657},
  doi        = {10.1145/1806651.1806657},
  language   = {en},
  urldate    = {2026-03-27},
  booktitle  = {Proceedings of the 2010 international symposium on {Memory} management},
  publisher  = {ACM},
  author     = {Nagarakatte, Santosh and Zhao, Jianzhou and Martin, Milo M.K. and Zdancewic, Steve},
  month      = jun,
  year       = {2010},
  pages      = {31--40}
}

@inproceedings{nagarakatte2009softbound,
  title     = {{SoftBound}: {Highly} compatible and complete spatial memory safety for {C}},
  booktitle = {Proceedings of the 30th {ACM} {SIGPLAN} {Conference} on {Programming} {Language} {Design} and {Implementation}},
  author    = {Nagarakatte, Santosh and Zhao, Jianzhou and Martin, Milo MK and Zdancewic, Steve},
  year      = {2009},
  pages     = {245--258}
}

@misc{nsa2023memory,
  author       = {{National Security Agency}},
  title        = {Software {Memory} {Safety}},
  year         = {2022},
  month        = nov,
  howpublished = {Cybersecurity Information Sheet,
                  \url{https://media.defense.gov/2022/Nov/10/2003112742/-1/-1/0/CSI_SOFTWARE_MEMORY_SAFETY.PDF}}
}

@inproceedings{nethercote2007valgrind,
  address    = {San Diego California USA},
  title      = {Valgrind: a framework for heavyweight dynamic binary instrumentation},
  isbn       = {978-1-59593-633-2},
  shorttitle = {Valgrind},
  url        = {https://dl.acm.org/doi/10.1145/1250734.1250746},
  doi        = {10.1145/1250734.1250746},
  language   = {en},
  urldate    = {2026-03-27},
  booktitle  = {Proceedings of the 28th {ACM} {SIGPLAN} {Conference} on {Programming} {Language} {Design} and {Implementation}},
  publisher  = {ACM},
  author     = {Nethercote, Nicholas and Seward, Julian},
  month      = jun,
  year       = {2007},
  pages      = {89--100}
}

@inproceedings{bruening2011practical,
  author    = {Bruening, Derek and Zhao, Qin},
  title     = {Practical Memory Checking with {Dr. Memory}},
  booktitle = {2011 International Symposium on Code Generation and Optimization},
  publisher = {IEEE},
  year      = {2011},
  pages     = {213--223},
  doi       = {10.1109/CGO.2011.5764689}
}

@inproceedings{duck2018effectivesan,
  address   = {Philadelphia, PA, USA},
  title     = {{EffectiveSan}: Type and Memory Error Detection using Dynamically Typed {C/C++}},
  url       = {https://dl.acm.org/doi/10.1145/3192366.3192388},
  doi       = {10.1145/3192366.3192388},
  booktitle = {Proceedings of the 39th {ACM} {SIGPLAN} Conference on Programming Language Design and Implementation},
  publisher = {ACM},
  author    = {Duck, Gregory J. and Yap, Roland H. C.},
  month     = jun,
  year      = {2018},
  pages     = {181--195}
}

@misc{oncd2024back,
  author       = {{Office of the National Cyber Director}},
  title        = {Back to the {Building} {Blocks}: {A} {Path} {Toward} {Secure} and {Measurable} {Software}},
  year         = {2024},
  month        = feb,
  howpublished = {The White House,
                  \url{https://bidenwhitehouse.archives.gov/wp-content/uploads/2024/02/Final-ONCD-Technical-Report.pdf}}
}

@misc{perens1993efence,
  author       = {Perens, Bruce},
  title        = {Electric {Fence} {Malloc} {Debugger}},
  year         = {1993},
  howpublished = {\url{https://github.com/kallisti5/ElectricFence}}
}

@inproceedings{serebryany2012addresssanitizer,
  title     = {{A}ddress{S}anitizer: {A} fast address sanity checker},
  booktitle = {2012 {USENIX} annual technical conference ({USENIX} {ATC} 12)},
  author    = {Serebryany, Konstantin and Bruening, Derek and Potapenko, Alexander and Vyukov, Dmitriy},
  year      = {2012},
  pages     = {309--318}
}

@misc{serebryany2018memory,
  author       = {Serebryany, Kostya and Stepanov, Evgenii and
                  Shlyapnikov, Aleksey and Tsyrklevich, Vlad and
                  Vyukov, Dmitry},
  title        = {Memory {Tagging} and how it improves {C}/{C}++ memory safety},
  year         = {2018},
  month        = feb,
  howpublished = {arXiv:1802.09517}
}

@inproceedings{shahriar2010classification,
  title     = {Classification of static analysis-based buffer overflow detectors},
  booktitle = {2010 {Fourth} {International} {Conference} on {Secure} {Software} {Integration} and {Reliability} {Improvement} {Companion}},
  publisher = {IEEE},
  author    = {Shahriar, Hossain and Zulkernine, Mohammad},
  year      = {2010},
  pages     = {94--101}
}

@inproceedings{stepanov2015memorysanitizer,
  address    = {San Francisco, CA, USA},
  title      = {{MemorySanitizer}: {Fast} detector of uninitialized memory use in \{{C}\}++},
  isbn       = {978-1-4799-8161-8},
  shorttitle = {{MemorySanitizer}},
  url        = {http://ieeexplore.ieee.org/document/7054186/},
  doi        = {10.1109/CGO.2015.7054186},
  urldate    = {2026-03-27},
  booktitle  = {2015 {IEEE}/{ACM} {International} {Symposium} on {Code} {Generation} and {Optimization} ({CGO})},
  publisher  = {IEEE},
  author     = {Stepanov, Evgeniy and Serebryany, Konstantin},
  month      = feb,
  year       = {2015},
  pages      = {46--55}
}

@article{vanoorschot2023memory,
  title      = {Memory {Errors} and {Memory} {Safety}: {C} as a {Case} {Study}},
  volume     = {21},
  issn       = {1558-4046},
  shorttitle = {Memory {Errors} and {Memory} {Safety}},
  url        = {https://ieeexplore.ieee.org/abstract/document/10102611},
  doi        = {10.1109/MSEC.2023.3236542},
  number     = {2},
  urldate    = {2026-03-17},
  journal    = {IEEE Security \& Privacy},
  author     = {van Oorschot, Paul C.},
  month      = mar,
  year       = {2023},
  pages      = {70--76}
}

@inproceedings{vintila2025evaluating,
  title     = {Evaluating the {Effectiveness} of {Memory} {Safety} {Sanitizers}},
  issn      = {2375-1207},
  url       = {https://ieeexplore.ieee.org/abstract/document/11023435},
  doi       = {10.1109/SP61157.2025.00088},
  urldate   = {2026-03-17},
  booktitle = {2025 {IEEE} {Symposium} on {Security} and {Privacy} ({SP})},
  author    = {Vintila, Emanuel Q. and Zieris, Philipp and Horsch, Julian},
  month     = may,
  year      = {2025},
  pages     = {774--792}
}

@inproceedings{watson2015cheri,
  address    = {San Jose, CA},
  title      = {{CHERI}: {A} {Hybrid} {Capability}-{System} {Architecture} for {Scalable} {Software} {Compartmentalization}},
  isbn       = {978-1-4673-6949-7},
  shorttitle = {{CHERI}},
  url        = {https://ieeexplore.ieee.org/document/7163016/},
  doi        = {10.1109/SP.2015.9},
  urldate    = {2026-03-27},
  booktitle  = {2015 {IEEE} {Symposium} on {Security} and {Privacy}},
  publisher  = {IEEE},
  author     = {Watson, Robert N.M. and Woodruff, Jonathan and Neumann, Peter G. and Moore, Simon W. and Anderson, Jonathan and Chisnall, David and Dave, Nirav and Davis, Brooks and Gudka, Khilan and Laurie, Ben and Murdoch, Steven J. and Norton, Robert and Roe, Michael and Son, Stacey and Vadera, Munraj},
  month      = may,
  year       = {2015},
  pages      = {20--37}
}

@article{woodruff2019cheri,
  title      = {{CHERI} {Concentrate}: {Practical} {Compressed} {Capabilities}},
  volume     = {68},
  copyright  = {https://ieeexplore.ieee.org/Xplorehelp/downloads/license-information/IEEE.html},
  issn       = {0018-9340, 1557-9956, 2326-3814},
  shorttitle = {{CHERI} {Concentrate}},
  url        = {https://ieeexplore.ieee.org/document/8703061/},
  doi        = {10.1109/TC.2019.2914037},
  number     = {10},
  urldate    = {2026-03-27},
  journal    = {IEEE Transactions on Computers},
  author     = {Woodruff, Jonathan and Joannou, Alexandre and Xia, Hongyan and Fox, Anthony and Norton, Robert M. and Chisnall, David and Davis, Brooks and Gudka, Khilan and Filardo, Nathaniel W. and Markettos, A. Theodore and Roe, Michael and Neumann, Peter G. and Watson, Robert N. M. and Moore, Simon W.},
  month      = oct,
  year       = {2019},
  pages      = {1455--1469}
}

@article{dion2023libpatch,
  title   = {Libpatch - {Dynamic} {Patching} of {Binaries} in {Userspace}},
  journal = {Software: Practice and Experience},
  author  = {Dion, Olivier and Desnoyers, Mathieu and Nassiri, Mohammad and Dagenais, Michel},
  year    = {2023}
}

@inproceedings{ainsworth2020markus,
  address    = {San Francisco, CA, USA},
  title      = {{MarkUs}: {Drop}-in use-after-free prevention for low-level languages},
  copyright  = {https://ieeexplore.ieee.org/Xplorehelp/downloads/license-information/IEEE.html},
  isbn       = {978-1-7281-3497-0},
  shorttitle = {{MarkUs}},
  url        = {https://ieeexplore.ieee.org/document/9152661/},
  doi        = {10.1109/SP40000.2020.00058},
  urldate    = {2026-03-27},
  booktitle  = {2020 {IEEE} {Symposium} on {Security} and {Privacy} ({SP})},
  publisher  = {IEEE},
  author     = {Ainsworth, Sam and Jones, Timothy M.},
  month      = may,
  year       = {2020},
  pages      = {578--591}
}

@inproceedings{silvestro2017freeguard,
  address    = {Dallas Texas USA},
  title      = {{FreeGuard}: {A} {Faster} {Secure} {Heap} {Allocator}},
  isbn       = {978-1-4503-4946-8},
  shorttitle = {{FreeGuard}},
  url        = {https://dl.acm.org/doi/10.1145/3133956.3133957},
  doi        = {10.1145/3133956.3133957},
  language   = {en},
  urldate    = {2026-03-27},
  booktitle  = {Proceedings of the 2017 {ACM} {SIGSAC} {Conference} on {Computer} and {Communications} {Security}},
  publisher  = {ACM},
  author     = {Silvestro, Sam and Liu, Hongyu and Crosser, Corey and Lin, Zhiqiang and Liu, Tongping},
  month      = oct,
  year       = {2017},
  pages      = {2389--2403}
}

@article{pereira2021characterizing,
  title    = {Characterizing {Buffer} {Overflow} {Vulnerabilities} in {Large} {C}/{C}++ {Projects}},
  volume   = {9},
  issn     = {2169-3536},
  url      = {https://ieeexplore.ieee.org/abstract/document/9576064},
  doi      = {10.1109/ACCESS.2021.3120349},
  urldate  = {2026-03-17},
  journal  = {IEEE Access},
  author   = {Pereira, José D’Abruzzo and Ivaki, Naghmeh and Vieira, Marco},
  year     = {2021},
  pages    = {142879--142892}
}

@techreport{black2018juliet,
  author      = {Black, Paul E.},
  title       = {Juliet 1.3 Test Suite: Changes From 1.2},
  institution = {National Institute of Standards and Technology},
  type        = {Technical Note},
  number      = {NIST TN 1995},
  year        = {2018},
  doi         = {10.6028/NIST.TN.1995}
}

@manual{adler2023pigz,
  author  = {Adler, Mark},
  title   = {{pigz}: A Parallel Implementation of {gzip}},
  year    = {2023},
  month   = aug,
  url     = {https://zlib.net/pigz/pigz.pdf},
  urldate = {2026-09-11}
}

\end{document}